%% file: 0main.tex
\documentclass[10pt,journal]{IEEEtran}
\IEEEoverridecommandlockouts

\usepackage{amsmath,amssymb,amsfonts}
\usepackage{textcomp}
\usepackage{epsfig}
\usepackage{listings}
\usepackage{diagbox}
\usepackage{epstopdf}

\usepackage{url}
\usepackage{cite}
\usepackage{multirow}
\usepackage{dsfont}
\usepackage{graphicx}
\usepackage[hang]{subfigure}
\usepackage{array}
\usepackage{mathtools}
\usepackage{algorithmic}

\usepackage{balance}
\usepackage{seqsplit}
\usepackage{pifont}
\usepackage{listings}
\usepackage{tikz}
\usepackage{fancyhdr}
\usepackage{enumitem}

\usepackage{soul}

\usepackage{color}

\newcommand{\blue}[1]{\textcolor{blue}{\bf #1}}
\newcommand{\nop}[1]{}

\begin{document}

\title{On Manually Reverse Engineering Communication Protocols of Linux Based IoT Systems}

\author{Kaizheng Liu,
        Ming Yang,
        Zhen Ling,
        Huaiyu Yan,
        Yue Zhang,
        Xinwen Fu,
        and Wei Zhao,~\IEEEmembership{Fellow,~IEEE}
        \vspace{-0.3cm}
\IEEEcompsocitemizethanks{\IEEEcompsocthanksitem K. Liu, M. Yang, Z. Ling and H. Yan are with the School of Computer Science
and Engineering, Southeast University, Nanjing 211189, China. 
(e-mail: kzliu@seu.edu.cn; yangming2002@seu.edu.cn; zhenling@seu.edu.cn; huaiyu\_yan@seu.edu.cn)
\IEEEcompsocthanksitem Y. Zhang is with the College of Information Science and Technology, Jinan University, Guangzhou 510632, China. 
(e-mail: zyueinfosec@gmail.com)
\IEEEcompsocthanksitem X. Fu is with the Department of Computer Science, University of Massachusetts Lowell, Lowell, MA 01854, USA.
(e-mail: xinwenfu@cs.uml.edu)
\IEEEcompsocthanksitem W. Zhao is with American University of Sharjah, PO Box 26666, Sharjah, UAE. 
(e-mail: weizhao@aus.edu)}}


\maketitle

\begin{abstract}

IoT security and privacy has raised grave concerns. Efforts have been made to design tools to identify and understand vulnerabilities of IoT systems. Most of the existing protocol security analysis techniques rely on a well understanding of the underlying communication protocols. In this paper, we systematically present the first manual reverse engineering framework for discovering communication protocols of embedded Linux based IoT systems. We have successfully applied our framework to reverse engineer a number of IoT systems. As an example, we present a detailed use of the framework reverse-engineering the WeMo smart plug communication protocol by extracting the firmware from the flash, performing static and dynamic analysis of the firmware and analyzing network traffic. The discovered protocol exposes severe design flaws that allow attackers to control or deny the service of victim plugs. Our manual reverse engineering framework is generic and can be applied to both read-only and writable Embedded Linux filesystems. 
\end{abstract}




\def\UrlBreaks{\do\A\do\B\do\C\do\D\do\E\do\F\do\G\do\H\do\I\do\J
\do\K\do\L\do\M\do\N\do\O\do\P\do\Q\do\R\do\S\do\T\do\U\do\V
\do\W\do\X\do\Y\do\Z\do\[\do\\\do\]\do\^\do\_\do\`\do\a\do\b
\do\c\do\d\do\e\do\f\do\g\do\h\do\i\do\j\do\k\do\l\do\m\do\n
\do\o\do\p\do\q\do\r\do\s\do\t\do\u\do\v\do\w\do\x\do\y\do\z
\do\.\do\@\do\\\do\/\do\!\do\_\do\|\do\;\do\>\do\]\do\)\do\,
\do\?\do\'\do+\do\=\do\#}

\input{1Introduction}
\input{2Background}

\input{3Overview}
\input{4Design.tex}
\input{5Casestudy}
\input{6Discussion.tex}
\input{7RelatedWork.tex}
\input{8Conclusion}

\balance
\bibliographystyle{IEEEtran}
\bibliography{wemo}

\end{document}

%% file: 1Introduction.tex
\section{Introduction}
\label{sec::Introduction}

Security of IoT products has received increasing scrutiny as IoT is being pervasively deployed
\cite{EuroSP::eyal::2016, Security::celik::2018,ZJY+::IoT::2019, EDBT::bertino::2016,JXM+::AWS::20}. 
For example, smart plugs and routers may be fully controlled by buffer overflow or command injection attacks \cite{dsp-w215, CCTVbufferoverflow, dir-815}. Security vulnerabilities also exist in popular IoT platforms such as AWS IoT \cite{ZJY+::IoT::2019,JXM+::AWS::20}. 

Efforts have been made to design tools to identify and understand vulnerabilities of IoT systems.
For example, Chen et al. \cite{NDSS::chen::2018} proposed an automatic fuzzing framework to find the memory corruption vulnerabilities caused by the software and firmware of IoT devices.
Given a well-formed protocol, formal and heuristic methods could be used to study security and identify the vulnerabilities of the protocol \cite{ACSAC::kim::2017, ADHoc::benjamin::2016, CNS::mohsin::2016, Globecom::ling::2017, IoJT::ling::2017}. For example, Kim et al. \cite{ACSAC::kim::2017} used formal symbolic modeling to automatically analyze the frequently-used IoT protocols, such as CoAP and MQTT.
Only when these protocols have been formally verified (mathematically proved) could they be considered as secure.
However, the challenge of automatic protocol verification relies on a well understood protocol. 

In this paper, we propose a framework of manually reverse engineering communication protocols of embedded Linux based IoT systems so that automation techniques can be applied over the discovered protocols for vulnerability discovery and security analysis. We focused on the embedded Linux based IoT system given its popularity. we find most IoT devices (more than 71\%) are installed with Linux, according to the Eclipse IoT developer survey \cite{eclipse2018}. Our framework adopts network traffic analysis and static analysis and dynamic analysis of the app and device firmware to understand specific details such as fields of the communication. 
Our manual reverse engineering framework works as follows: (i) Obtaining the app and firmware of the device; (ii) Collecting network traffic generated by the device and app with testbeds; (iii) Defeating traffic protection by using the man-in-the-middle (MITM) proxy, static analysis and dynamic debugging to defeat traffic encryption and obfuscation; (iv) Discovering the communication protocol through traffic analysis, static analysis and dynamic analysis of the app and firmware. 

We have applied our framework and reverse engineered a number of IoT systems including smart plugs, IP cameras and air quality monitoring sensors. As an example, this paper presents a detailed case study of the popular WeMo smart plug from Belkin.
The plug system involves three parts: smart plugs, smartphones, and two cloud servers. A smartphone can communicate with a smart plug via the cloud servers. The cloud servers distribute keys to the smartphone and smart plug, and authenticate them based on the distributed keys. Once the communication protocol of the smart plug is discovered, we are able to identify a serious design flaw that allows two attacks: (i) A malicious software smartphone bot could be used to control victim plugs; (ii) A fake smart plug can pretend to be a real one and kick the real one offline.

\textbf{Contribution}: Major contributions of this paper can be summarized as follows:
(i) We are the first to systematically propose a framework to manually reverse engineer communication protocols of IoT systems.
(ii) We have applied this framework to successfully reverse engineer a number of IoT systems. As an example, this paper presents a complete protocol analysis of the WeMo smart plug and identifies severe design flaws that allow attackers to control victim plugs and deny the service of victim plugs. We also briefly discuss how we apply the framework to a few other IoT systems. (iii) Our communication protocol reverse engineering framework is generic and can be applied to both read-only and writable Linux filesystems. We collected the firmware of 514 popular IoT devices on the market and showed that our framework is applicable to them.

\textbf{Road map}: The rest of the paper is organized as follows. In Section \ref{sec::backgroud}, we briefly introduce background knowledge. In Section \ref{sec::Analysis_Method}, we present our communication protocol reverse engineering framework. In Section \ref{sec::Reverse}, we present a case study of the WeMo smart plug using the proposed framework. 
In Section \ref{sec::Evaluation}, we discuss the generality and limitations of our framework. Related work is presented in Section \ref{sec::RelatedWork} and we conclude this paper in Section \ref{sec::Conclusion}.

%% file: 2Background.tex
\section{Background}
\label{sec::backgroud}

In this section, we present a brief introduction to the architecture of an IoT system and terms used in this paper.

\subsection{Architecture of an IoT System}
\label{subsubsec:iotframe}

\begin{figure}
    \centering
    \includegraphics[width=.45\textwidth]{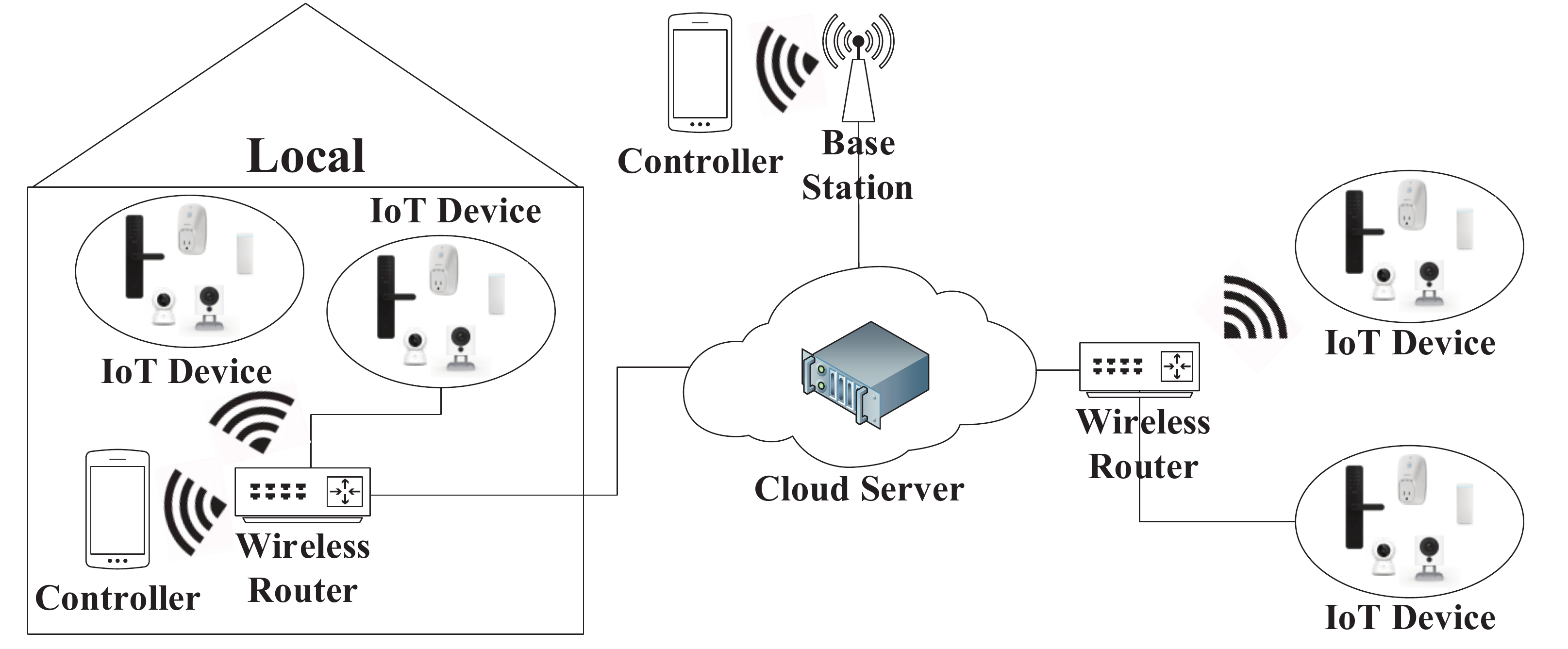}
    \vspace{-0.1cm}
    \caption{A Simplified Architecture of an IoT system}
    \vspace{-0.5cm}
    \label{Architecture_img}
\end{figure}

Fig. \ref{Architecture_img} shows a typical IoT communication system based on our experiments and previous researches \cite{ESORICS::yao::2019, Globecom::ling::2017, IoJT::ling::2017, Security::zhou::2019}. The system consists of three components, an IoT device, a controller and a cloud server. The IoT device implements specific functionalities, such as medical monitoring and electrical control. The controller, such as a smartphone app, is used to control the IoT device. The cloud server is used to relay messages between the controller and the IoT device. The cloud server may provide other services including device management, data storage and analysis.
For a smart plug system, the smart plug is the IoT device while the smartphone app is the controller. When the controller and IoT device are located in the same network, the smart plug’s official app could be used to directly communicate and control the plug through WiFi. If the controller and IoT device are in different networks, a cloud server could be adopted to transmit the message between the controller and the IoT device so as to traverse the NAT (Network Address Translation).

\subsection{Communication Protocols and Terms}
\label{subsec:protocolprocess}
An IoT communication system may realize complicated communication protocols and various functionalities. We have identified four common phases of an IoT communication protocol, including paring, binding, authentication and controlling \cite{Globecom::ling::2017, mass::chao::2018}, which are crucial for the overall system security. 
(i) Pairing: To bootstrap and configure an IoT device, a user often needs to connect a controller (e.g. an app on a smartphone) to the IoT device via various communication venues. For example, the IoT device can work as a WiFi access point (AP) so that the controller can connect to it. The controller can also connect to the IoT device via Bluetooth. We denote this connecting process as pairing. This is relevant to security since the pairing process may be under malicious sniffing and anyone may get access to the IoT device, particularly in the cases that the device is deployed in public.
(ii) Binding: When pairing is completed, a binding mechanism is often employed so that the cloud server can associate the controller and IoT device, and relay messages between them.
(iii) Authentication: The controller, device and cloud server often need to authenticate each other to defeat various threats and abuses. 
(iv) Controlling: After authentication, the controller can take control of the IoT device via a cloud server or a local network.

%% file: 3Overview.tex
\section{Framework of Manually Reverse Engineering IoT Communication Protocols}
\label{sec::Analysis_Method}

In this section, we will present the assumption about capabilities of security analysts, and our manual reverse engineering framework.

\vspace{-0.1cm}
\subsection{Capabilities of Security Analyst}
\label{subsec:threatmodel}

We adopt the term ``security analyst" to refer to those who would use our framework to reverse engineer third party IoT products. 
We make the following assumptions about the capabilities of the security analyst:
(i) The analyst can obtain IoT devices of interest and the controller applications, as well as set up a testbed. Without loss of generality, an Android app is used as an example of the controller. 
Most IoT vendors provide both Android and iOS apps, and the communication protocol for the Android app and iOS app is the same. Therefore, whether the Android app or iOS app is analyzed, the analyst can extract the complete communication protocol.
(ii) We focus on IoT devices that use the popular and open source embedded Linux based Operation System (OS).   

\begin{figure}
    \centering
    \includegraphics[width=0.4\textwidth]{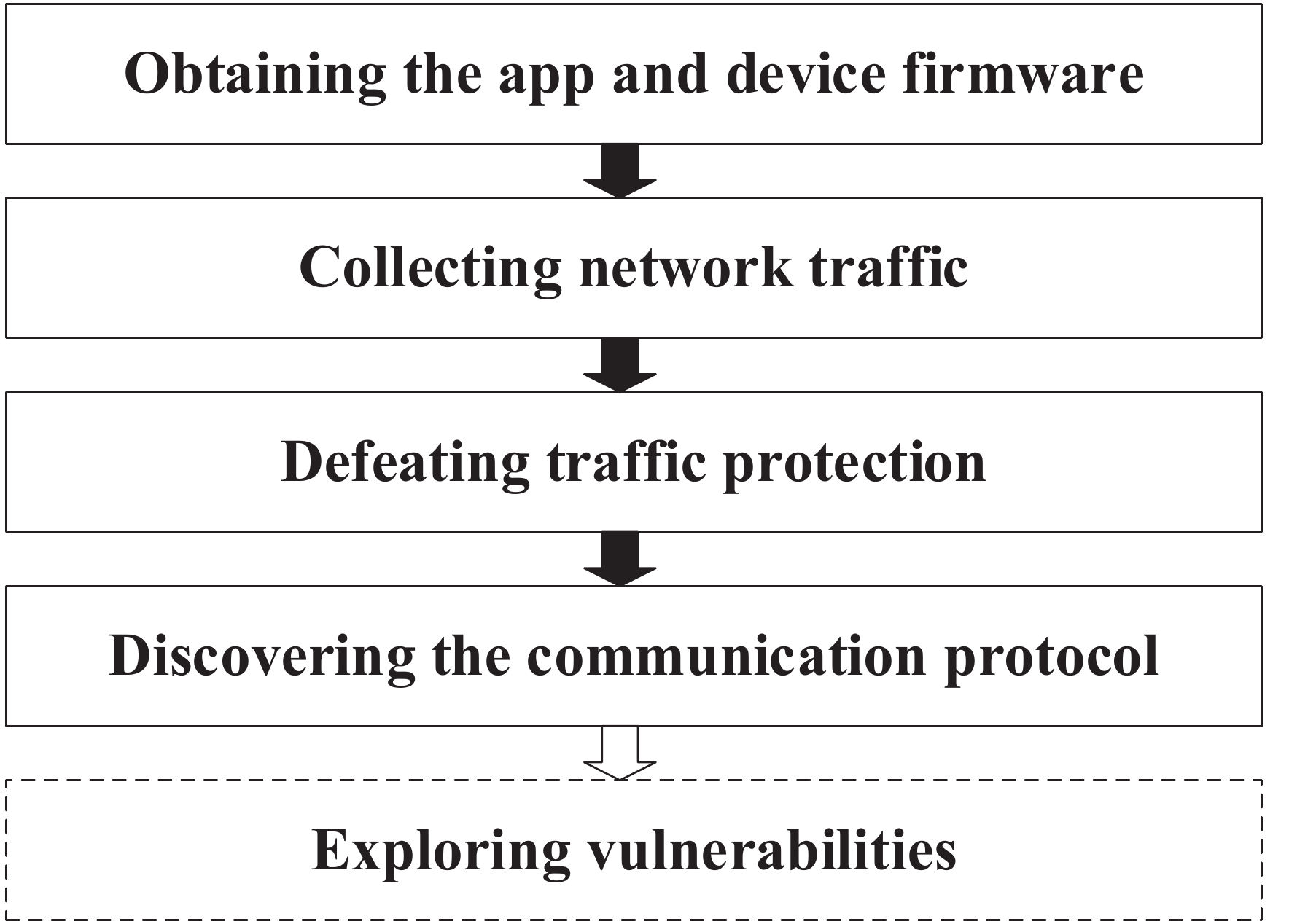}
    \vspace{-0.1cm}
    \caption{Workflow of communication protocol reverse engineering framework}
    \label{flow_analyze_method}
    \vspace{-0.3cm}
\end{figure}

\vspace{-0.1cm}
\subsection{Overview}


Fig. \ref{flow_analyze_method} illustrates the workflow of our manual reverse engineering framework: obtaining the app and device firmware, collecting network traffic, defeating traffic protection, and discovering the communication protocol. 

\begin{enumerate}
\item {Obtaining the  app and device firmware}: 
The app is often free and can be downloaded from Google Play (or Apple App Store). The device firmware may be obtained from the manufacturer's website, over-the-air (OTA) update process \cite{shavit2007firmware} (i.e., firmware update process) or reading the flash chip as discussed later in this section. The first two approaches are straightforward. However, they may not be always available.

\item {Collecting network traffic}: 
In this step, we particularly want to collect network traffic and understand security related phases of the IoT communication protocols. During the pairing process, the IoT device may work as a WiFi Access Point (AP) and the controller connects to this AP. A sniffer is needed to dump the pairing traffic. After pairing, the device and controller will connect to the Internet through a router/switch/AP. For simplicity, we will use AP to refer to router/switch/AP. To intercept the traffic after pairing, we set up our own APs. The controller and IoT device connect to our APs and communicate with each other through either the local network or Internet. The traffic of interest can be collected from these APs.

\item {Defeating traffic protection}:
Some vendors may adopt TLS/SSL encryption or obfuscation to protect the communication. The analyst can defeat the TLS/SSL encryption with a MITM proxy. Obfuscation algorithms can be disclosed through static analysis and dynamic debugging of the app and firmware.

\item {Discovering the communication protocol}: 
Through the combination of traffic analysis, static analysis and dynamic analysis of the app and firmware, the communication protocol can be discovered. Based on the discovered communication protocol, the analyst may use either heuristic methods or formal methods to find vulnerabilities of the protocol. In this paper, we use heuristic methods to demonstrate the feasibility of the reverse engineering approach.
\end{enumerate}

%% file: 4Design.tex


\subsection{Obtaining the App and Device Firmware}
\label{subsec:obtaining}

The app is often free and can be downloaded from Google Play. However, it can be a challenge to extract the firmware from the flash chip, which often involves the following steps. First, we take apart the physical device and identify the device's flash chip model (e.g. NOR flash, NAND flash) and packaging type (e.g. small-outline package (SOP), quad flat package (QFP), ball grid array (BGA)). The information can be found on the surface of the chip or the case of the IoT device. With such information, we can determine which type of surface-mount packaging is applied to the device's flash chip accordingly. For example, if the flash uses SOP that often exposes the flash pins, we can connect Bus Pirate \cite{BusPirate} to the corresponding pins via a test clip and an adapter in order to read the firmware image from the flash.
However, with a particular packaging technology, for example, BGA, a flash chip may not expose its pins. In such a case, we may de-solder the flash chip by using an Surface Mount Technology (SMT) rework station \cite{Oh::REFlash::Blackhat2014}. After obtaining the flash chip, a flash engineering programmer like StarProg-F \cite{StarProg} may be used to read the firmware image from the flash. 
 
\subsection{Collecting Network Traffic}
\label{subsec:sinff}

A wireless network card supporting the monitor mode can be used as a sniffer to dump WiFi traffic such as the pairing traffic.
To build our own AP or a wireless router, we install Hostapd \cite{hostapd} on a computer with a wireless network card supporting the AP mode.
The computer is also equipped with an Ethernet card connecting to the Internet.
Some IoT devices only support Ethernet. In such a case, we equip the computer with a second Ethernet card connecting to such an IoT device.
Therefore, the computer can intercept passing traffic from the IoT device or controller.
 
\vspace{-0.1cm}
\subsection{Defeating Traffic Protection}
\label{subsec:dealingtraffic}

We now discuss how to defeat encryption and obfuscation which are used to protect traffic from the app and IoT device. 

\subsubsection{Encryption}
\label{subsubsec:dealingencrypt}
Network traffic can be encrypted by TLS/SSL. To decrypt the traffic, a MITM transparent proxy is installed in front of the smartphone (i.e., controller) or IoT device. The proxy is used to relay or manipulate the traffic between the device and remote server, or the traffic between the smartphone and remote server.
With proper configuration, the MITM proxy can decrypt the passing traffic.
Specifically, we use an open source tool ``mitmproxy'' \cite{mitmproxy} as our MITM proxy. 

We now show how to replace the target root certificate
issued by a trusted certificate authority (CA) or a self-signed private root certificate with the forged root certificate on a controller. Take Android as an example. From our empirical analysis, the certificate can be located in three places: (i) The trusted CA certificate is stored in ``/system/etc/security" as an individual file \cite{androidCertificate}. 
In this case, we can just add the forged root certificate to the Android system.
(ii) The private root certificate can be packaged as a file in an app. In this case, we use APKTool \cite{apktool2019} to unpack the APK package and replace the original certificate with the forged root certificate. We then recompile and sign the APK \cite{signapk}. (iii) The private root certificate can also be hard-coded in the format of a string in the app code. In this case, we decompile the original app into smali code, identify and replace the original hard-coded root certificate, and finally generate a new app.

We now discuss how to replace the original root certificate with the forged root certificate on an IoT device. This case is more complicated.
(i) We first search the root certificate in the filesystem of the obtained firmware.
The original certificate can be a standalone file or hard-coded in a binary file. The certificate often has a set of features. For example, if the certificate is encoded in Privacy Enhanced Mail (PEM) \cite{rfc::spki::1999} format, it contains a header (``-\,-\,-\,-\,-BEGIN CERTIFICATE-\,-\,-\,-\,-''). Therefore, we can locate the certificate by searching the header.
(ii) Once we locate the root certificate, we need to identify which type of filesystem is used by the firmware so that a specific replacement method can be applied. An open source tool named Binwalk \cite{binwalk} is introduced to identify the filesystem type, either a writable filesystem, such as {\em JFFS2} and {\em UBIFS} or a read-only filesystem, such as {\em SquashFS} and {\em CramFS}.
For a read-only filesystem, the replacement cannot be made directly since
modification is not allowed. We can re-flash a customized firmware with the forged root certificate into the device. 
We may need to generate the Cyclical Redundancy Check (CRC) and append it to the customized firmware to pass the chip's integrity check. 
For a writable filesystem, there are two ways to replace the original certificate: (a) If we can get into the console of the IoT device system, for example, by using universal asynchronous receiver-transmitter (UART), and locate file transfer tools like a ftp client, we can replace the original certificate directly via file transfer tools.
(b) We can replace the original certificate directly by mounting the writable filesystem segmented from the firmware onto a Linux computer.
We then re-package a new firmware with the modified filesystem and flash the new firmware into the device.  

There are two ways to flash a modified firmware with the forged root certificate into an IoT device.
(i) We can flash the firmware back using Bus Pirate or a flash engineering programmer. If the flash chip is de-soldered for reading the firmware \cite{Oh::REFlash::Blackhat2014}, we need to re-solder it back to the circuit board. (ii) We can also flash the firmware back to the chip via the firmware upgrading interface like the over-the-air (OTA) interface. 

\subsubsection{Obfuscation}
An IoT system may protect its traffic by obfuscation. 
Traffic obfuscation is used to make communications more complicated. Unlike encryption, obfuscation does not require a key to encrypt or decrypt the traffic \cite{obfuscation}.
Static analysis and dynamic analysis may be adopted to counter traffic obfuscation. 

To de-obfuscate traffic from a controller, for example, an Android app, we first need to understand how the obfuscated traffic is generated and then write a de-obfuscation algorithm. To this end, we first need to check if the app is packed. For a packed app, we can unpack it \cite{ESORICS::dexhunter::2015, RAID::appspear::2015, ICSE::lei::2017}. Then we can extract the smali code using Apktool. We analyze the workflow of the traffic obfuscation algorithm by reading the extracted smali code. We may use Smali2Java \cite{smali2java2019} to decompile the smali code into the Java format for easy understanding.  
We can also dynamically debug the smali code by using Android SDK and Android Studio \cite{studio2017android} as follows: (i) We add a new field ``android:debuggable=true'' in {the tag} of Android manifest file ``application'' to enable debugging. (ii) We locate the function of the entry activity, ``onCreate'', and add a line of smali code at the beginning of this function as shown in List \ref{wait_for_debug} to make the app wait for the debug signal after being started. (iii) We repackage the modified APK and install it in the smartphone. 
(iv) Now once we start the app, 
we can use Android Studio to add break points and monitor the functions of interest.

{\scriptsize
\begin{lstlisting}[caption={Add waitForDebugger function to entry activity of app},label={wait_for_debug}, breaklines=true]
a=0;// # virtual methods
a=0;// .method protected onCreate(Landroid/os/Bundle;)V
a=0;//     invoke-static {}, Landroid/os/Debug;->waitForDebugger()V
\end{lstlisting}
}

\begin{figure*}[htbp]
\centering
    \begin{minipage}[t]{0.23\linewidth}
    \centering
    \includegraphics[height=3.8cm]{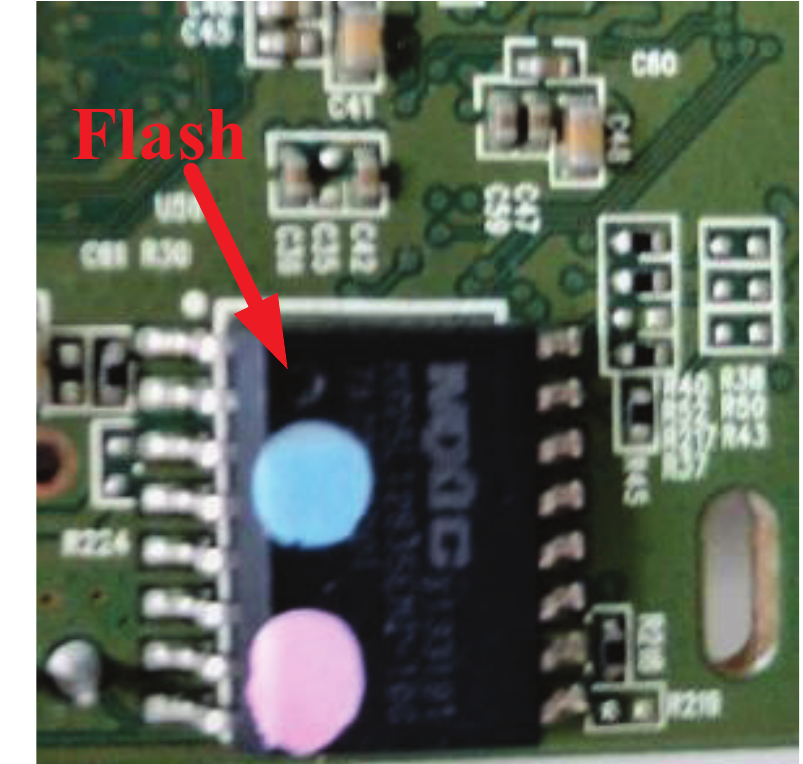}
    \caption{Flash of WeMo plug}
    \vspace{-0.2cm}
    \label{flash}
    \end{minipage}
    \begin{minipage}[t]{0.23\linewidth}
    \centering
    \includegraphics[height=3.8cm]{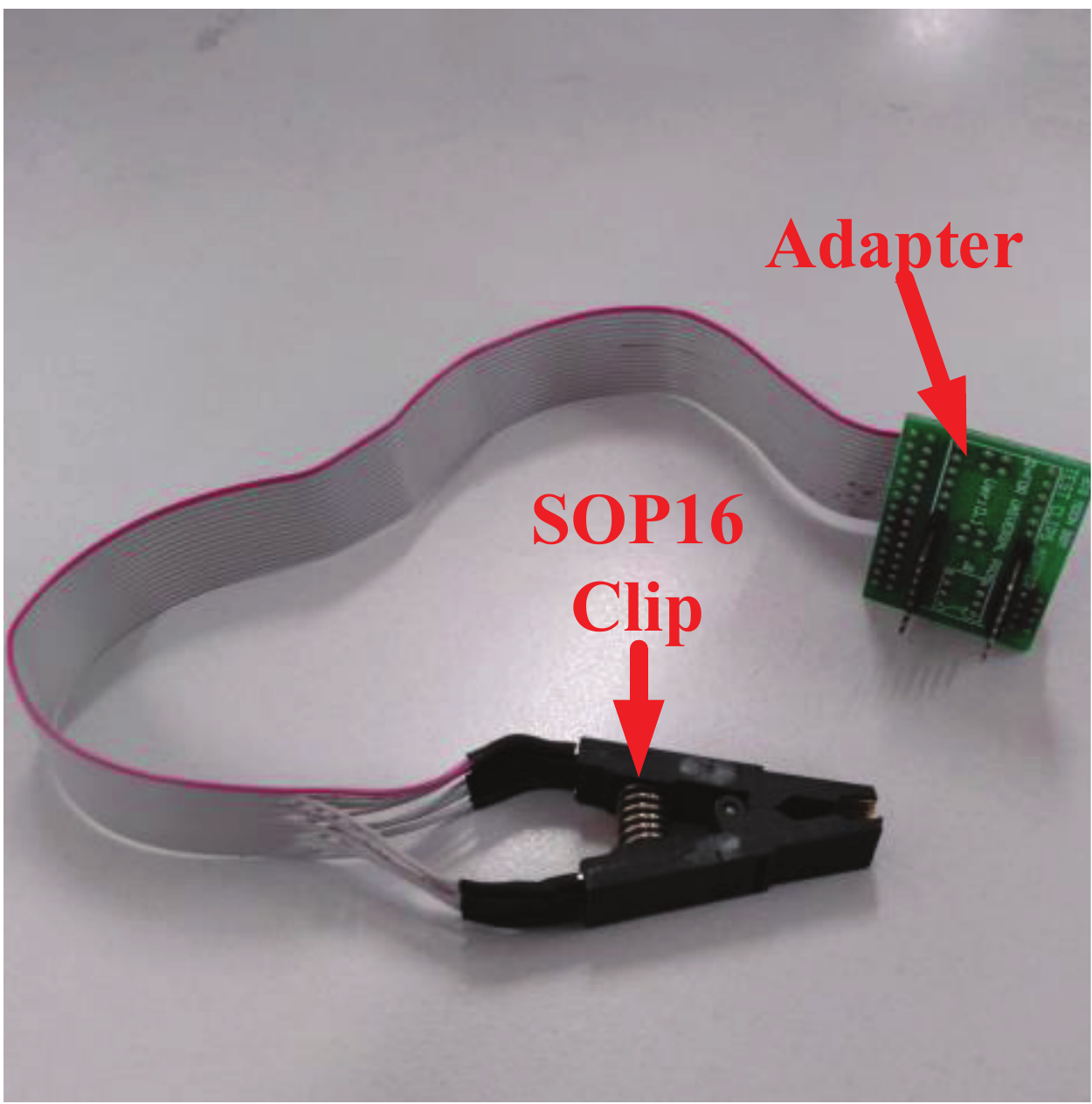}
    \caption{Adapter and SOP16 Clip}
    \vspace{-0.2cm}
    \label{adpter}
    \end{minipage}
    \begin{minipage}[t]{0.23\linewidth}
    \centering
    \includegraphics[height=3.8cm]{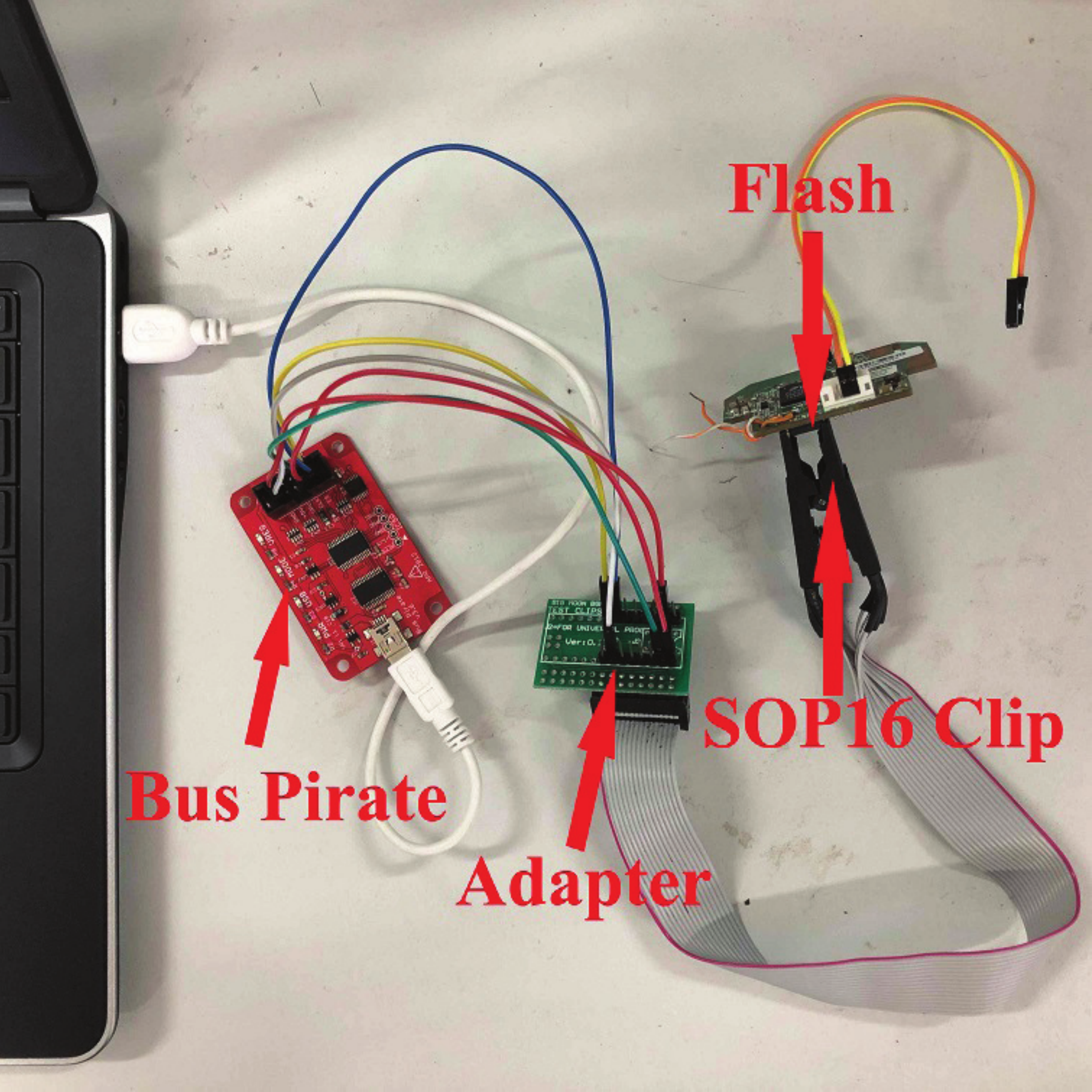}
    \caption{Reading Flash by Bus Pirate}
    \vspace{-0.2cm}
    \label{BusPirate}
    \end{minipage}
    \begin{minipage}[t]{0.23\linewidth}
    \centering
    \includegraphics[height=3.8cm]{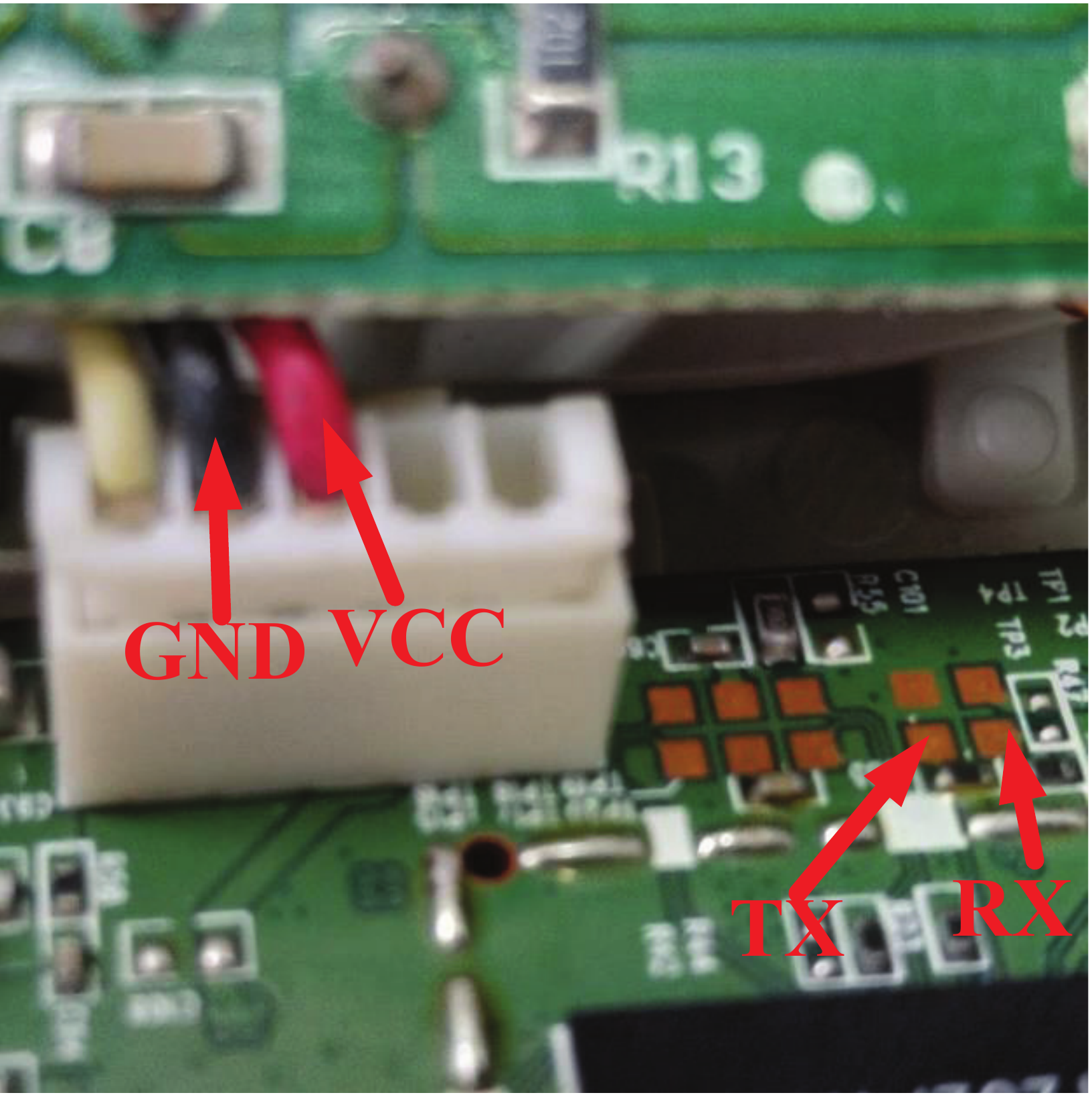}
    \caption{UART of WeMo plug}
    \vspace{-0.2cm}
    \label{uart_wemo}
    \end{minipage}
\end{figure*}      

\begin{figure*}[htbp]
\begin{center}
    \begin{minipage}[t]{0.3\linewidth}
    \centering
    \includegraphics[width=0.74\linewidth]{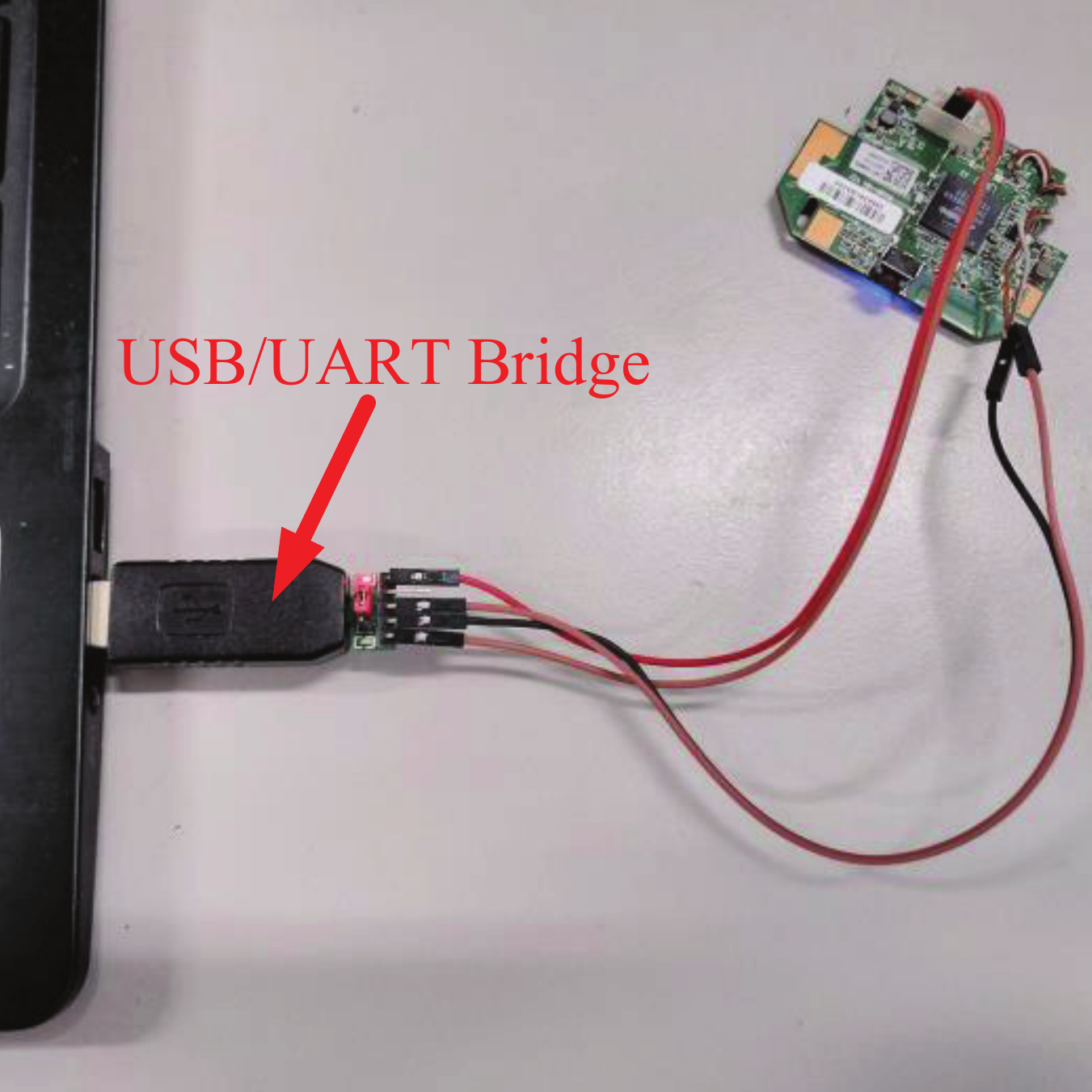}
    \caption{Open a console of WeMo plug system by UART}
   
    \label{UART}
    \vspace{-0.6cm}
    \end{minipage}
    \begin{minipage}[t]{0.25\linewidth}
    \centering
    \includegraphics[width=0.8\linewidth]{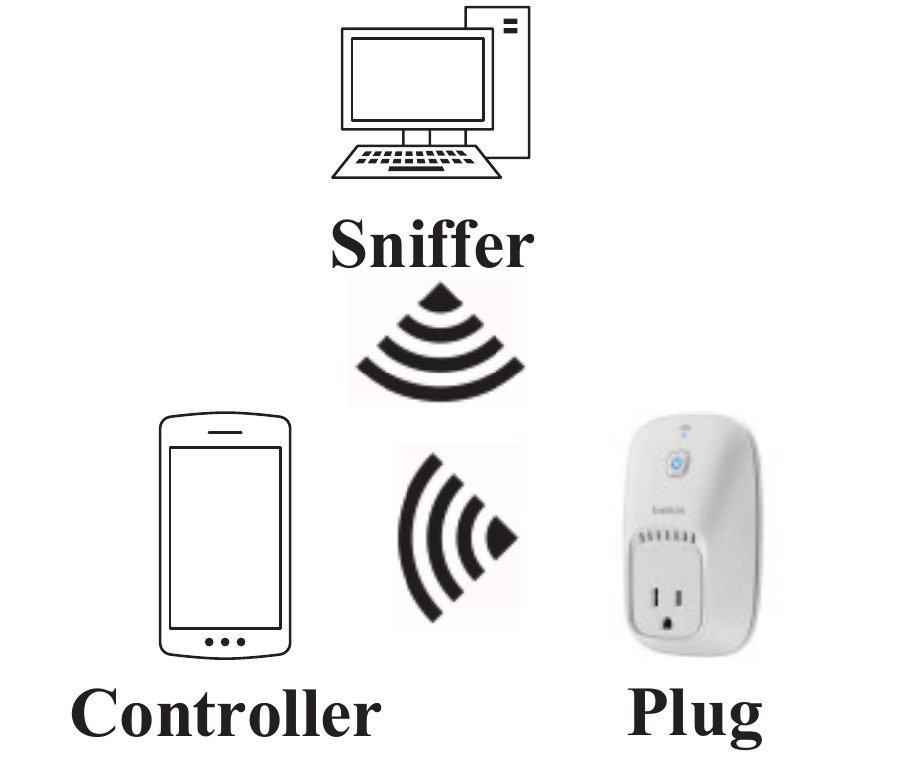}
    \caption{Pairing traffic collection}
    \label{wemo_mitm_pair}
    \vspace{-0.6cm}
    \end{minipage}
    \begin{minipage}[t]{0.4\linewidth}
    \centering
    \includegraphics[width=1\linewidth]{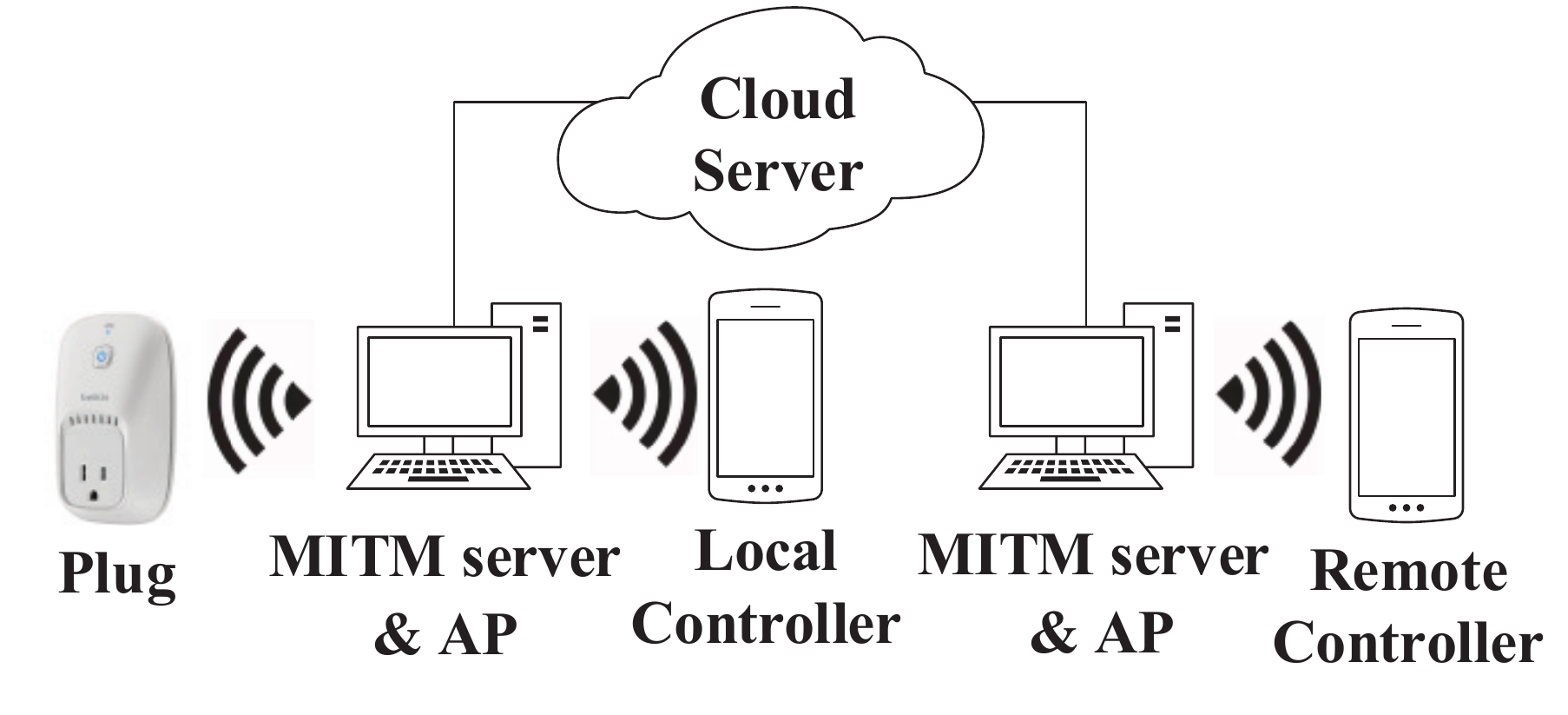}
    \caption{NAT traffic collection}
    \label{wemo_mitm_non_pair}
    \vspace{-0.6cm}
    \end{minipage}
\end{center}
\end{figure*}  

However, the method above will fail when Java Native Interface (JNI) is applied. To address this issue, we introduce IDA pro \cite{idapro}, a multi-platform tool that offers both static and dynamic analysis functionalities.
(i) We first enable USB debugging on the tested smartphone.
(ii) We copy the  binary file of IDA pro, ``android\_server'', to the smartphone  and run it via the Android Debug Bridge (adb) \cite{google::adb}.
(iii) We map a port on the computer to a port on the smartphone so that they can communicate with each other. 
(iv) We run the app in debug mode, and start the IDA pro client on the computer. The smartphone then forwards the debug log to the computer via the configured port. 
 
We now discuss {\em how to de-obfuscate traffic from an IoT device}. We  need to identify the algorithm that obfuscates the messages and write a de-obfuscation algorithm. The obfuscation algorithm is usually stored in a particular binary file. Therefore, the first step is to identify this file in the firmware. We compare the information from the analysis of dumped network traffic with the IoT device's runtime system log. If a match is discovered, the file can then be identified. To obtain the log, we first need to obtain the console of the IoT device system. If we can locate the UART port on the board of IoT device, 
we can connect it to the debugging computer using a UART-to-USB bridge with a correct baud rate. 
Otherwise, we can embed a backdoor such as telnet into the IoT device firmware and update the device with the new firmware. A telnet app is often hidden in an IoT device maybe for the purpose of debugging by the manufacturer and can be utilized too. We can then log in the IoT device system through the backdoor from the debugging computer.
The log can then be shown in the console of the computer after the IoT device starts. For example, we are often interested in the design flaws in authentication of the controller and IoT device. Hence we perform the authentication phase repeatedly and compare the ports used in each process in the runtime system log with the port of intercepted obfuscated traffic. If the ports match, we find the target binary file. Afterwards, we extract the binary file with Binwalk from the firmware of the IoT device as discussed in Section \ref{subsec:obtaining}. 

Once obtaining the binary file, in order to obtain the obfuscation algorithm, we can analyze it as follows: (i) We can perform static analysis to disassemble the binary file with IDA pro.  
(ii) We can also dynamically analyze it on the IoT device using binary instrumentation \cite{ISPASS::laurenzano::2010} by inserting additional code into the executable binary file to observe or modify the behavior of the binary file. Binary instrumentation allows us to trace functions of interest, and follow the workflow of the inputs and outputs. To use binary instrumentation, we need to modify the firmware with the method proposed in Section \ref{subsubsec:dealingencrypt}.  
(iii) We can also use the GDB client and GDBserver \cite{stallman2002debugging} to remotely debug the binary program of the IoT device from a computer. We first need to cross compile the GDBserver and embed the GDBserver into the firmware of the target IoT device and run the GDB client in our debugging computer. By configuring the IP address and port of the GDBserver, we can use the GDB client to dynamically debug the target binary file and identify the traffic obfuscation algorithm.

\vspace{-0.1cm}
\subsection{Discovering the Communication Protocols}
\label{subsec:protocolanalysis}
Through traffic analysis, we may understand the basics of the communication protocols. For details like encrypted/obfuscated fields, we perform the following procedure to understand them. 
(i) We may measure the entropy of the bytes of the traffic to determine whether the command or data is created with cryptographic operations such as encryption and hash. High entropy beyond a threshold indicates the data is encrypted or hashed.
(ii) We may also search cryptographic APIs within the firmware to determine if encryption is used and also identify cryptographic functions that are used. At the controller side, the developers may encrypt or hash the application layer data using cryptographic APIs of Android SDK or C/C++ libraries. 
We can use dynamic analysis tools (e.g., ``Xposed'' \cite{xposed} and ``Frida'' \cite{Frida:2020}) to hook the frequently-used cryptographic APIs \cite{NDSS::zuo::2016}. Once a specific cryptographic function is called, the information of this function is recorded. Therefore, we know which function is used. At the device side, we can employ static data flow analysis to identify a cryptographic function \cite{arzt2017static}.
(iii) Once we locate the target cryptographic function, we can 
obtain the original command or data and the key for the cryptographic function by dynamically debugging the binary file and analyzing the inputs of the target cryptographic function with the method introduced in Section \ref{subsec:dealingtraffic}. Specifically, we can use the ``Xposed'' and ``Frida'' at the controller side and use the GDB debugging tool at the device side, respectively.

\vspace{-0.1cm}
\subsection{Exploring Vulnerabilities}
With the discovered protocols, we can use heuristic methods or formal methods to perform security analysis of the IoT communication protocol and identify potential vulnerabilities. We find the vulnerabilities in the four phases of an IoT protocol (pairing, binding, authentication and controlling introduced in Section \ref{sec::backgroud}) often incur severe damages \cite{Globecom::ling::2017, IoJT::ling::2017}. The security analyst may focus on these four phases while performing vulnerability assessment of IoT systems.

\nop{\subsection{Reconstructing communication protocols}
\blue{Based on the previous analysis, the entire communication protocol is discovered. Then we can reconstruct the communication protocol which can be used for security analysis. Based on previous researches, we find the vulnerabilities in the four components of an IoT protocol (pairing, binding, authentication, and control introduced in Section \ref{sec::backgroud}) often incur severe damages \cite{Globecom::ling::2017, IoJT::ling::2017}. The security analyst can focus on these four components while performing a vulnerability assessment of IoT systems.}}

%% file: 5Casestudy.tex
\section{Case Study: Smart Plugs from Belkin WeMo}
\label{sec::Reverse}

The manual reverse engineering framework introduced in Section \ref{subsec:obtaining} is the result of our reverse engineering of a number IoT devices, including our previous research \cite{Globecom::ling::2017, IoJT::ling::2017}.
In this section, we present a case study of reverse engineering the WeMo smart plug using the framework and the discovered communication protocols. We will also introduce novel attacks against the plug based on the discovered protocols. 

\subsection{Reverse Engineering WeMo Smart Plug}
We present the workflow of reverse engineering the WeMo smart plug.

\subsubsection{Obtaining the app and device firmware}

The official app of the smart plug is free to download while the firmware is publicly unavailable. 
The flash chip of the smart plug is shown in Fig. \ref{flash} and it is packaged with SOP. As shown in Fig. \ref{BusPirate}, we can use Bus Pirate to read the firmware from the flash chip with a SOP16 clip and an adapter, which are shown in Fig. \ref{adpter}.

\subsubsection{Collecting network traffic}

A testbed is deployed to eavesdrop on the network traffic of interest. As shown in Fig. \ref{wemo_mitm_pair}, 
during the pairing phase, the smart plug works as an AP and we collect the pairing traffic with a sniffer.
We intercept the traffic between the smartphone, smart plug and cloud server by introducing two APs, as shown in Fig. \ref{wemo_mitm_non_pair}. 
\begin{figure}
      \centering
      \includegraphics[width=0.45\textwidth]{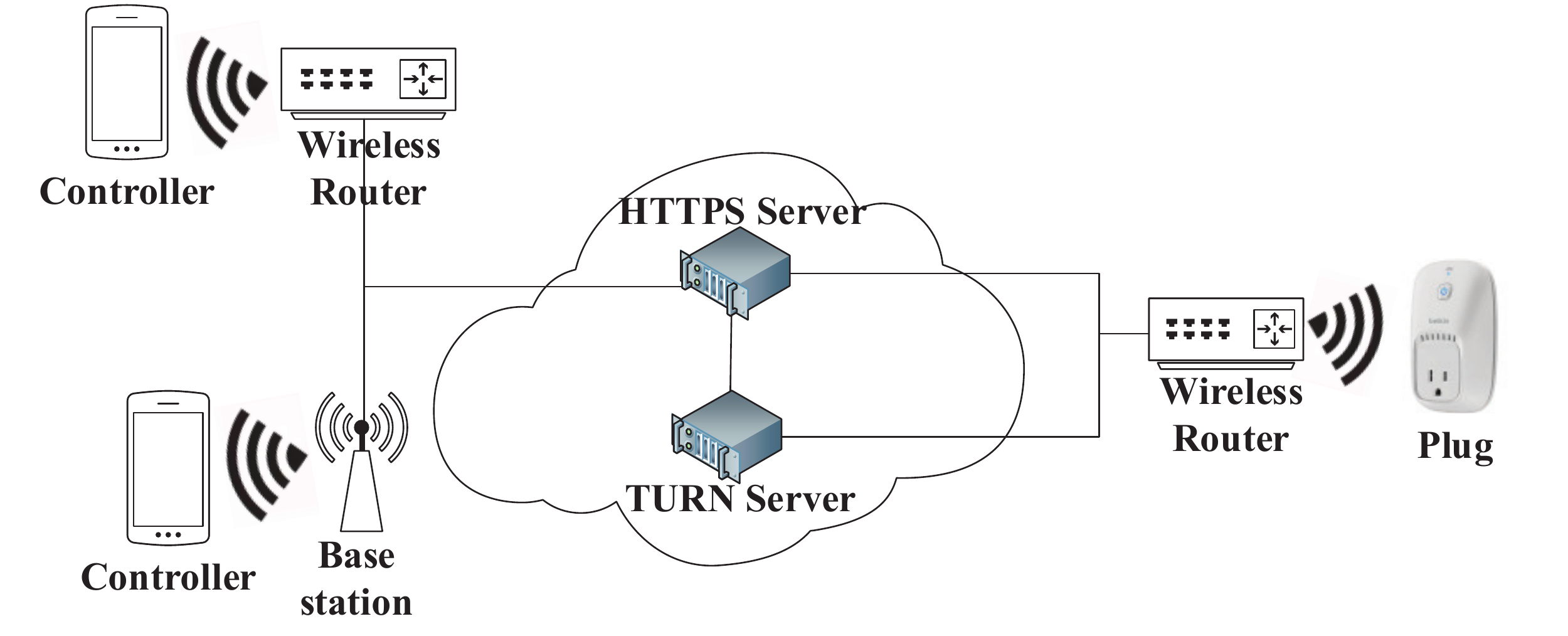}
      \vspace{-0.1cm}
      \caption{Architecture of WeMo plug system}
      \label{Architecture_wemo}
      \vspace{-0.6cm}
\end{figure}

\subsubsection{Defeating traffic protection}

The primary challenge of decrypting encrypted traffic is to replace the original certificate of the firmware and controller app with our forged one. (i) We first replace the certificate of the smartphone. We find that the original certificate is stored in ``/system/etc/security''. Therefore, on the smartphone, we can download the forged root certificate generated by the MITM proxy through a web browser and Android will prompt us to install the certificate.
(ii) We then replace the original CA certificate in the firmware of the smart plug. 
We find a UART port on the chip as shown in Fig. \ref{uart_wemo}, where the UART port has four pins, including TX, RX, GND and VCC. We use a UART-to-USB bridge to open a console of the smart plug's embedded Linux system, as shown in Fig. \ref{UART}, and find a ftp client in the system. We put the forged root certificate on a ftp server and download it to the plug system through the discovered ftp client, so as to replace the original CA certificate. The forged certificate will be preserved in the device even after the device reboots. This shows the plug's filesystem is writable. 
By using Binwalk, we find that the firmware actually contains a read-only {\em SquashFS} filesystem and a writable {\em JFFS2} filesystem.
The plug system implements a virtual filesystem, ``mini\_fo'', which merges the read-only {\em SquashFS} filesystem and the writable {\em JFFS2} filesystem. When a file is changed, the new file is written to the writable {\em JFFS2} filesystem while the read-only {\em SquashFS} filesystem still keeps the original file. 
(iii) After the certificate is successfully replaced, we can eavesdrop on connections with ``mitmproxy''.

\subsubsection{Discovering the communication  protocol}

We now present how to reverse engineer the smart plug's application layer protocol. Based on traffic analysis, we are able to identify strings that start with ``MESSAGE-INTEGRITY'' or ``Authorization'', but other fields of such strings are unreadable.
We find that these fields are generated with the HMAC-SHA1 algorithm \cite{krawczyk1997hmac} by using the methods in Section \ref{subsec:protocolanalysis}. These fields actually contain authentication materials, which are crucial for our security analysis. 



\vspace{-0.1cm}
\subsection{Communication Protocols of WeMo Smart Plug}
\label{subsec::Protocols}

We now present the discovered architecture of the WeMo smart plug system, and its communication protocols.

\begin{figure}
      \centering
      \includegraphics[width=0.4\textwidth]{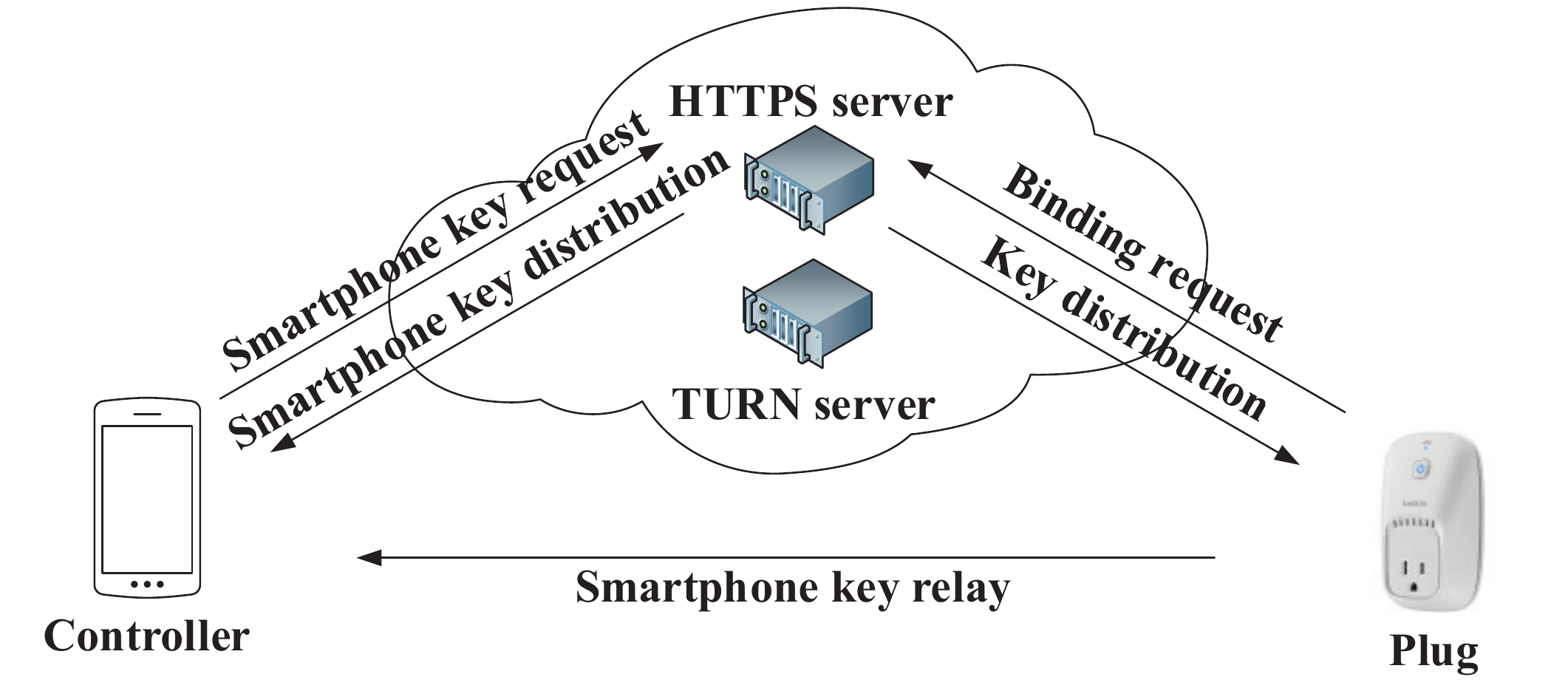}
      \vspace{-0.1cm}
      \caption{Binding phase}
      \label{Registration_phase_img}
      \vspace{-0.4cm}
\end{figure}
    
\subsubsection{Architecture of WeMo smart plug system}

The WeMo smart plug system contains three components: two cloud servers (a TURN server and a HTTPS server), smart plugs and smartphones. A smart plug and a smartphone can communicate with each other via the cloud servers, as shown in Fig. \ref{Architecture_wemo}. 
Since a smart plug is often behind a WiFi router using the NAT, the TURN \cite{perreault2010traversal} server is used to perform the NAT traversal for the plug so that a user on the Internet can send a command to the plug. The HTTPS server has three functionalities, including binding, authentication, and controlling (i.e., command relay and information update). 
 
\subsubsection{Pairing}
\label{subsubsec::pairing}
In the pairing phase, the plug works as an AP and the smartphone connects to it.
The smartphone sends a request to the plug to obtain basic information of the plug, such as the MAC address and serial number. After receiving such information, the smartphone 
sends the plug its identification (ID) and description, a timestamp $TS$, and the home AP's WiFi credentials entered by the user. Then the plug can access the Internet via the home AP.

\subsubsection{Binding}
\label{subsubsec:binding}
The smartphone and smart plug are bound to the HTTPS sever as shown in Fig. \ref{Registration_phase_img}.
The smart plug first sends the binding request, including MAC address, smartphone's ID and description of the plug, SSID and MAC address of WiFi, and timestamp $TS$ to the HTTPS server, which can now bind (associate) the particular plug and smartphone together on the basis of the received information.
Based on materials contained in the binding request, the HTTPS server produces two keys: the smart plug key and the smartphone key. The HTTPS server then sends these two keys to the smart plug. 
After obtaining the two keys, the smart plug sends the smartphone key to the smartphone via the local WiFi network. If the smart plug and smartphone are not in the same local network, the smartphone can obtain the smartphone key by sending a request to the HTTPS server that knows the particular smartphone is bound to the particular plug. 
The request also contains a message authentication code, as introduced below.

\begin{figure}
\centering
\includegraphics[width=.42\textwidth]{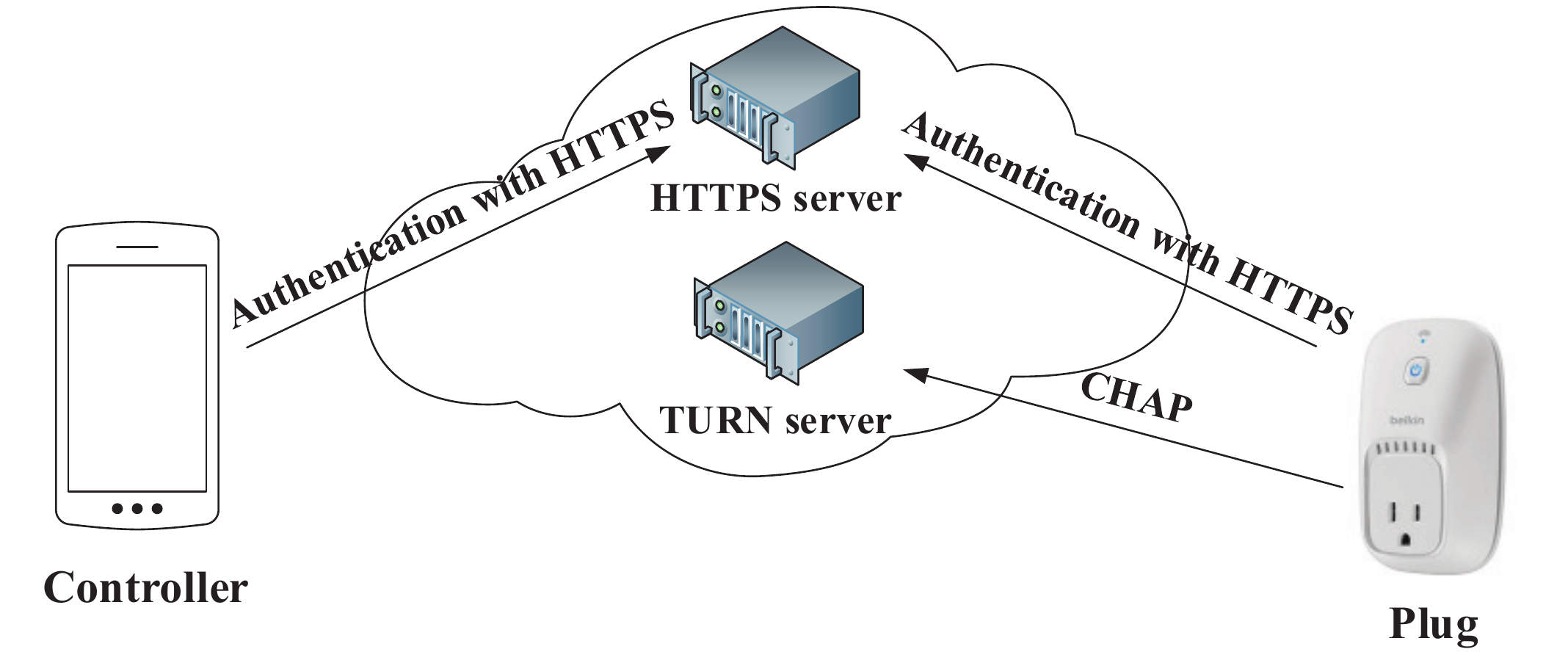}
\vspace{-0.1cm}
\caption{Authentication phase}
\label{Discovery_phase_img}
\vspace{-0.4cm}
\end{figure}
    
\subsubsection{Authentication}
\label{subsubsec:authenticate}   

Fig. \ref{Discovery_phase_img} summarizes the authentication phase. Within the local network, there is no authentication for a smartphone app to control the plug. When the smartphone and smart plug are not in the same local network, they need to communicate through the HTTPS server.
\nop{Take the plug as an example. }In each message from the plug to the HTPPS server, the HTTP message header includes an ``Authorization'' field, which contains authentication data. The authentication data is generated by the HMAC-SHA1 algorithm over the plug key and other shared information with the HTTPS server. 
The HTTPS server authenticates the smartphone in a similar way.
The TURN server obtains the smart plug key from the HTTPS server and authenticates the plug via the Challenge Handshake Authentication Protocol (CHAP) \cite{simpson1996ppp}.
  
\subsubsection{Remote Controlling} 

\begin{figure}
\centering
\includegraphics[width=.43\textwidth]{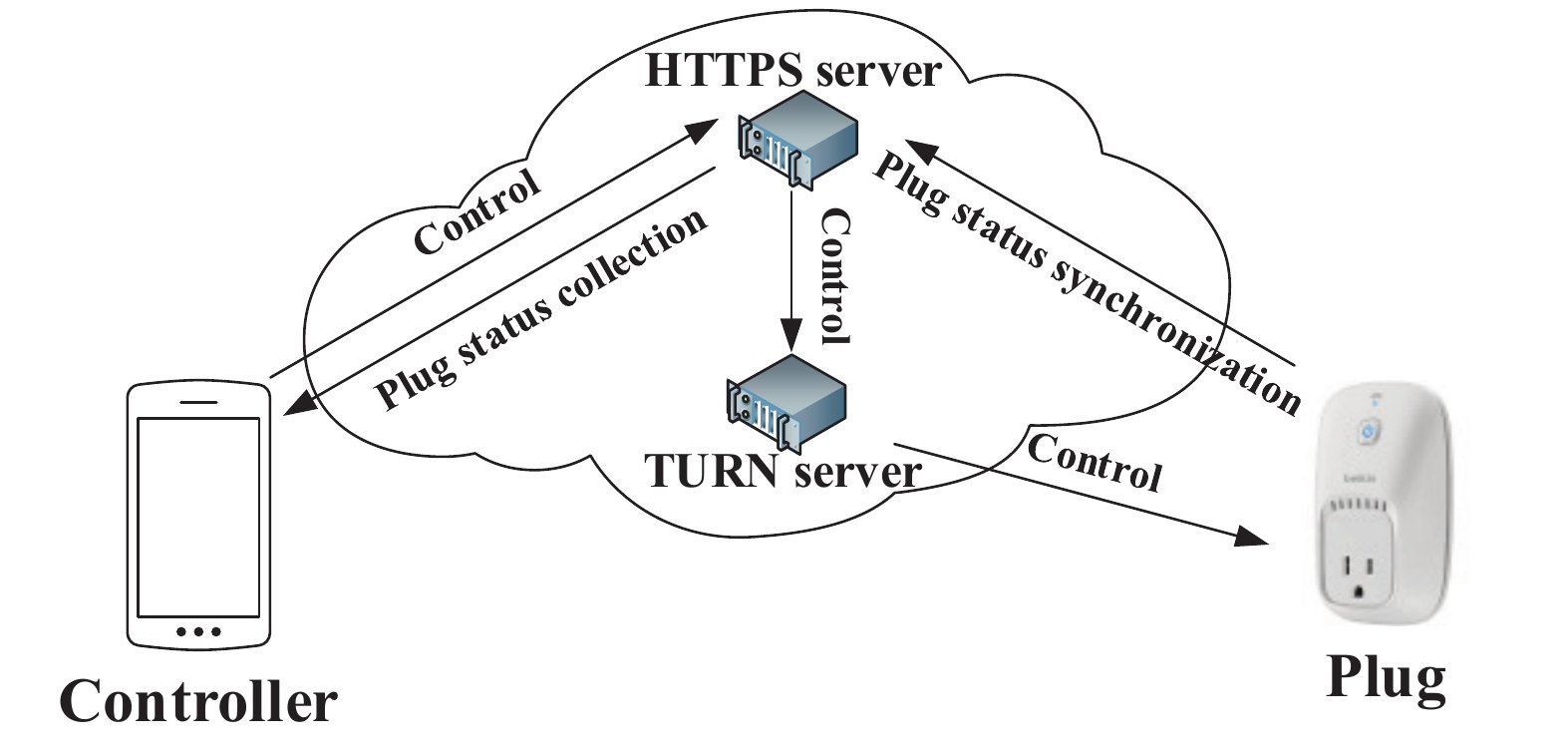}
\vspace{-0.1cm}
\caption{Remote controlling phase}
\label{Remote_control_img}
\vspace{-0.4cm}
\end{figure}

After authentication, the smartphone and smart plug can communicate with the cloud servers as illustrated in Fig. \ref{Remote_control_img}.
The smart plug periodically synchronizes its status with the HTTPS server. To remotely control the plug, the smartphone first obtains the status of the smart plug by sending a request to the HTTPS server. The status can be either {\em switch\_off} (integer ``0'') or {\em switch\_on} (integer ``1''). When the device is offline, the status is {\em unavailable} (integer ``3''). Then the smartphone can send control commands to switch on/off the smart plug via the the HTTPS server. The HTTPS server actually forwards the commands to the TURN server, which uses the NAT traversal to send the command through the wireless router to the plug.

\vspace{-0.1cm}
\subsection{Attacks against WeMo Plugs}
\label{sec::Attack}

Once the IoT communication protocols are discovered, we can now move forward with security analysis of pairing, binding, authentication and controlling phases introduced in Section \ref{subsec:protocolprocess}. We discovered two novel attacks against the WeMo smart plug: sharing attack and connection hijacking attack.
With the sharing attack, an attacker can remotely control a victim smart plug. The connection hijacking attack allows a DOS attack against a plug.
It is worth noting that all the experiments are conducted on the plugs that we purchase.

\subsubsection{Sharing attack}
\label{subsubsec:illegalsharing}

We first introduce the details of the binding phase, which involves two binding requests from the plug. The authorization value in the first binding request is ``dummy'', as the plug key is not derived yet. After receiving the first binding request, the HTTPS server sends back a temporary key. The authorization value in the second binding request from the plug is generated using the temporary key. After receiving the second binding request, the HTTPS server sends the plug key and smartphone key to the plug. 

To explore the smart plug resetting phase, we first press the reset button on a smart plug and then bind a new smartphone to the plug. We find now both the original and new smartphones can remotely control the plug. That is, the original and new smartphones now share the plug.
Through traffic analysis, we find the plug sends only one binding request, which is regarded as rebinding request, to the HTTPS server. The rebinding request contains a new field, ``reRegister''. The authorization value is generated using the original plug key. It can be inferred that the original plug key is not erased after resetting. 

We find that if we set the authorization value as ``dummy'' in the rebinding request to pretend that the smart plug loses its key, the HTTPS server will send the original plug key and a new smartphone key to the plug. Once the new smartphone obtains the new smartphone key, the smartphone can pass the authentication of the HTTPS server and access the plug.

Once we understand the plug sharing phase, we are able to bind a victim smart plug to a malicious smartphone. . 
The details of the sharing attack are introduced as follows.

\begin{enumerate}
\item  
To deploy the attack, the attacker needs to obtain the victim plug's MAC address and serial number, as well as the home AP's SSID and MAC address. 
One limitation of this attack is that the attacker has to use wardriving or other means to get the victim plug's MAC address and home AP's SSID and MAC address. 
In wardriving, the attacker drives around and performs wireless sniffing. Blocks of MAC addresses are allocated to every manufacturer (Belkin in our case), which can be obtained from the Internet. Therefore, the attacker will be able to identify Belkin smart plugs through wardriving.
We also find that a plug's serial number is predictable based on its MAC address. Therefore, the attacker can remotely attack the victim plug after obtaining the needed information.
  
\item The attacker can now implement a fake software smart plug that pretends to be the real one.
The fake plug sends a rebinding request with the authorization value ``dummy'' and fabricated smartphone information to the HTTPS server to get a temporary key. Once the plug receives the key, it resends a rebinding request with the authorization value that is generated by the temporary key, and then obtains the original plug key and a new smartphone key. 
  
\item The attacker now creates a fake software smartphone, which uses the new smartphone key and sends commands with correct authorization value to the HTTPS server. It is worth noting that the HTTPS server has already bound the victim plug and the fake smartphone together. In this way, the attacker can remotely control the target WeMo smart plug while the victim user cannot discover the attack for the sharing feature of the WeMo plug.
\end{enumerate}

\subsubsection{Connection hijacking attack}
\label{subsubsec:hijack}

Once obtaining the plug key through the sharing attack, a fake smart plug can pretend to be the real device so as to hijack the connection between the victim user and the real plug. 
The details of the attack process are presented as follows.
\begin{enumerate}
\item The attacker first creates a fake smart plug that pretends to be the real one and uses it to deploy the sharing attack in Section \ref{subsubsec:illegalsharing}. In this way the attacker obtains the victim smart plug key. 
\item Since the fake smart plug has the original smart plug key, the fake smart plug can perform the authentication process with the plug system's Traversal Using Relays around NAT (TURN) server to request a relay port, which is shared with the HTTPS server. Therefore, the HTTPS server knows that the fake plug uses that specific TURN server port.  
\item Now a control command from a victim smartphone is sent from the HTTPS server to the relay port of the fake plug on the TURN server. The command is relayed to the fake plug instead of the real one. The traffic from the smartphone is hijacked by the attacker, who denies the service of the victim smart plug as a matter of fact.
\end{enumerate}


\subsubsection{Discussion}
At the time of writing this paper, Belkin has added a security patch trying to defeat our sharing attack. With the patch, if the public source IP address of the rebinding request sent from a plug is changed, the HTTPS server will not send the original plug key, but generate a new smart plug key. 
Since the victim plug still keeps the old plug key, it will not be able to pass the authentication of the HTTPS server and TURN server, and cannot be controlled by a controller anymore. Therefore, our sharing attack becomes a DoS attack under the security patch. 
If a user wants to reuse the victim plug, he/she has to reset the plug.

%% file: 6Discussion.tex
\vspace{-0.1cm}
\section{Evaluation}
\label{sec::Evaluation}

\begin{table*}[!ht]
\begin{center}
\vspace{-0.2cm}
\caption{File system of IoT devices (CramFS, SquashFS and RomFS are read-only file systems and JFFS2 is a writable file system)}
\label{file_system_collection}
\vspace{-0.2cm}
\begin{tabular}{|c|c|c|c|c|c|c|c|c|}
\hline
\diagbox{File System}{Manufacturer}&Axis&Asmnet&D-Link&TP-Link&Netgaer&Netis&Asus&Total\\
\hline
CramFS&45&0&0&0&0&0&0&45\\
\hline
JFFS2&33&1&0&0&0&0&0&34\\
\hline
SquashFS&0&9&35&13&34&29&57&177\\
\hline
CramFS\&JFFS2&249&0&0&0&0&0&0&249\\
\hline
RomFS&6&3&0&0&0&0&0&9\\
\hline
Total&333&15&35&13&34&29&57&514\\
\hline
\end{tabular}
\vspace{-0.3cm}
\end{center}
\end{table*}

In this section, we evaluate the generality of our communication protocol reverse engineering framework, present our reverse engineering of a number of real-world IoT system, and discuss the limitations of the proposed framework.

\vspace{-0.1cm}
\subsection{Generality of our Manual Reverse Engineering Framework}

The most challenging part of reverse engineering an IoT device is firmware analysis. The firmware may be from different vendors with high customization.
Table \ref{file_system_collection} shows the mainstream manufacturers and the file systems used by their products. We collected 514 firmware from 7 vendors by crawling the Internet. By analyzing these firmware with binwalk, we can identify the file systems used in these firmware. For example, out of the 333 firmware published by Axis, 6 of them use RomFS, 45 use CramFS file system, 33 use JFFS2 file system and 249 use both CramFS and JFFS2 file systems.
The file system can be a read-only (e.g., CramFS, SquashFS or RomFS) or writable (e.g., JFFS2).
To reverse engineer these types of firmware, we often need to change the firmware, for example, embedding a fake CA certificate for mitmproxy or a GDBserver for debugging. 
We can perform such changes with approached introduced in Section \ref{subsec:dealingtraffic}. Therefore, we will be able to reserve engineer all the devices listed in Table \ref{file_system_collection} while the actual manual reverse engineering tasks may last long given the complexity.

\subsection{Reverse Engineering Real-world IoT Products}
\begin{figure}[h]
\centering

\subfigure[WeMo\textcolor{white}{+++} Plug]{%
\begin{minipage}[t]{0.21\linewidth}
\centering
\includegraphics[width=1.1\textwidth]{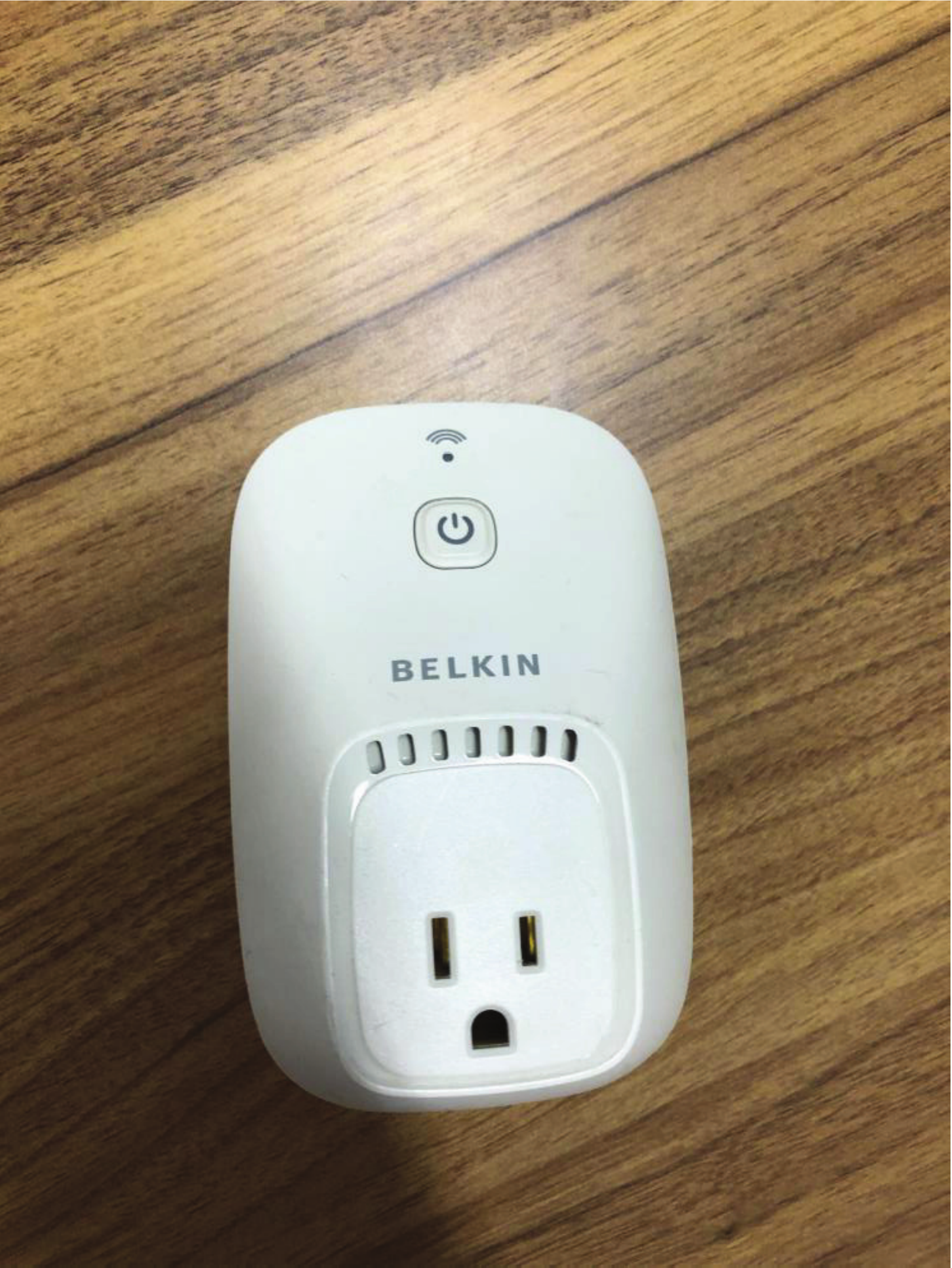}%
\end{minipage}
}
\subfigure[DLink\textcolor{white}{+} Camera]{%
\begin{minipage}[t]{0.21\linewidth}
\centering
\includegraphics[width=1.1\textwidth]{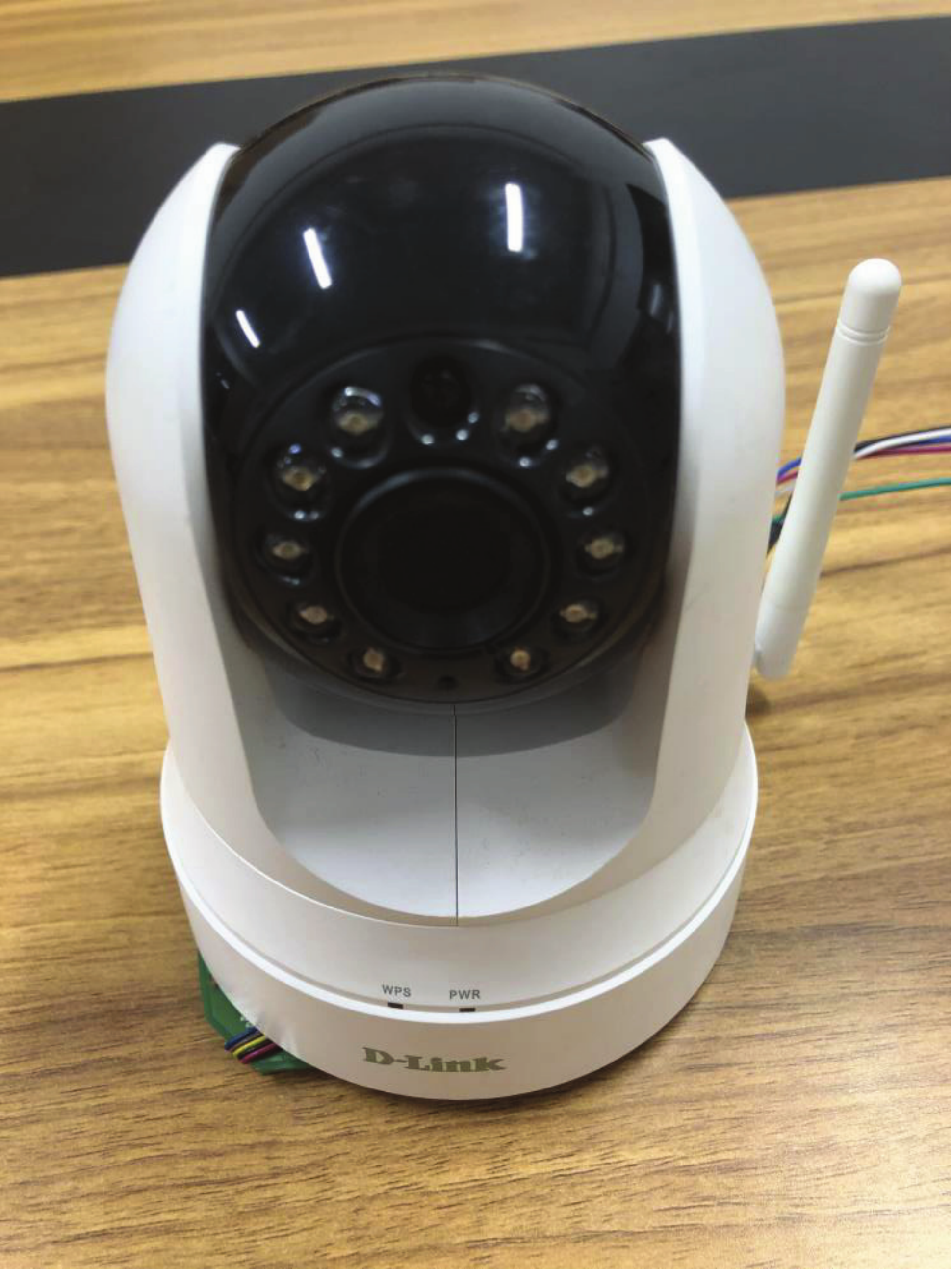}%
\end{minipage}
}
\subfigure[Haier Camera]{%
\begin{minipage}[t]{0.21\linewidth}
\centering
\includegraphics[width=1.1\textwidth]{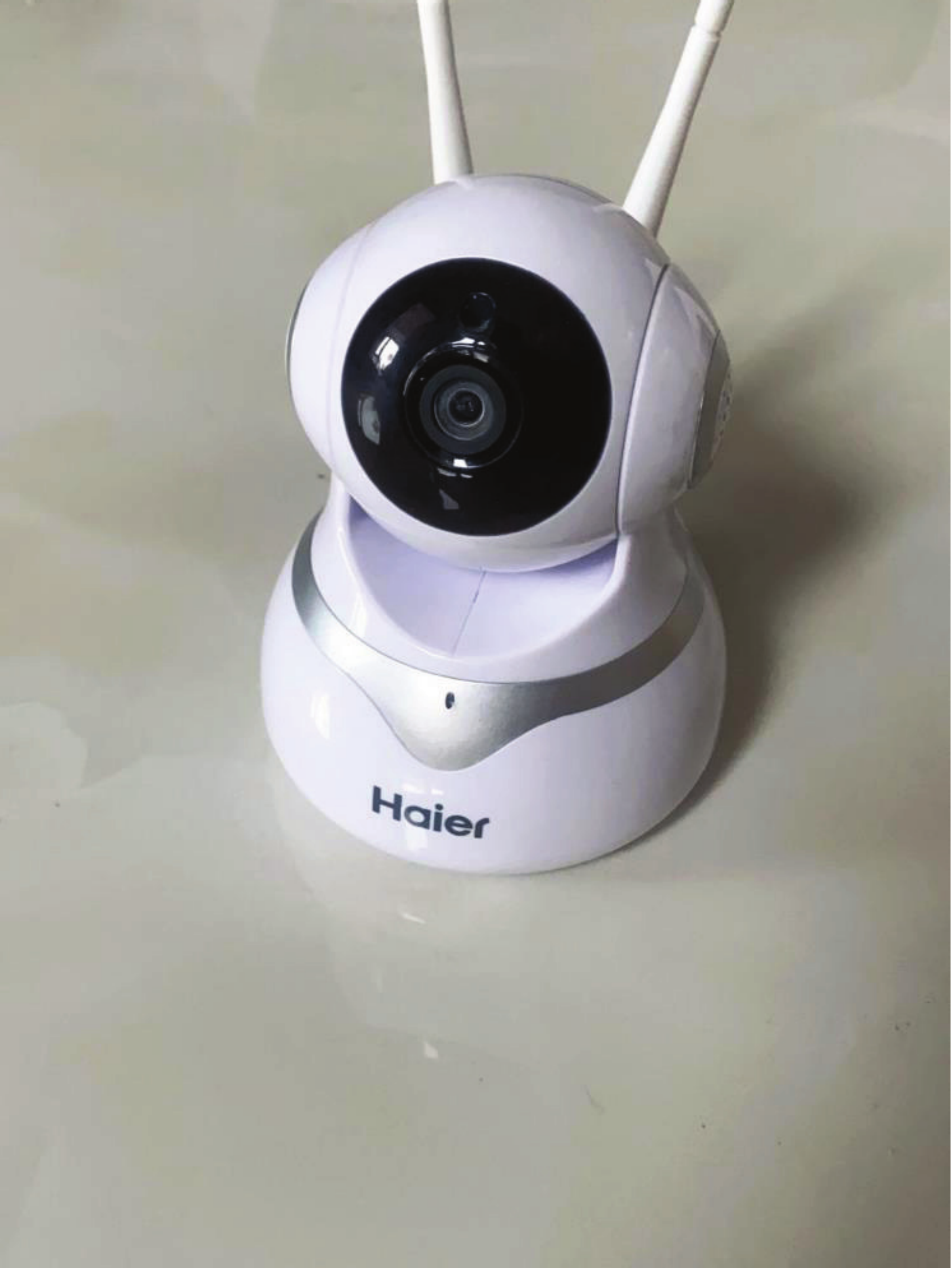}%
\end{minipage}
}
\subfigure[Xiongmai Camera]{%
\begin{minipage}[t]{0.21\linewidth}
\centering
\includegraphics[width=1.1\textwidth]{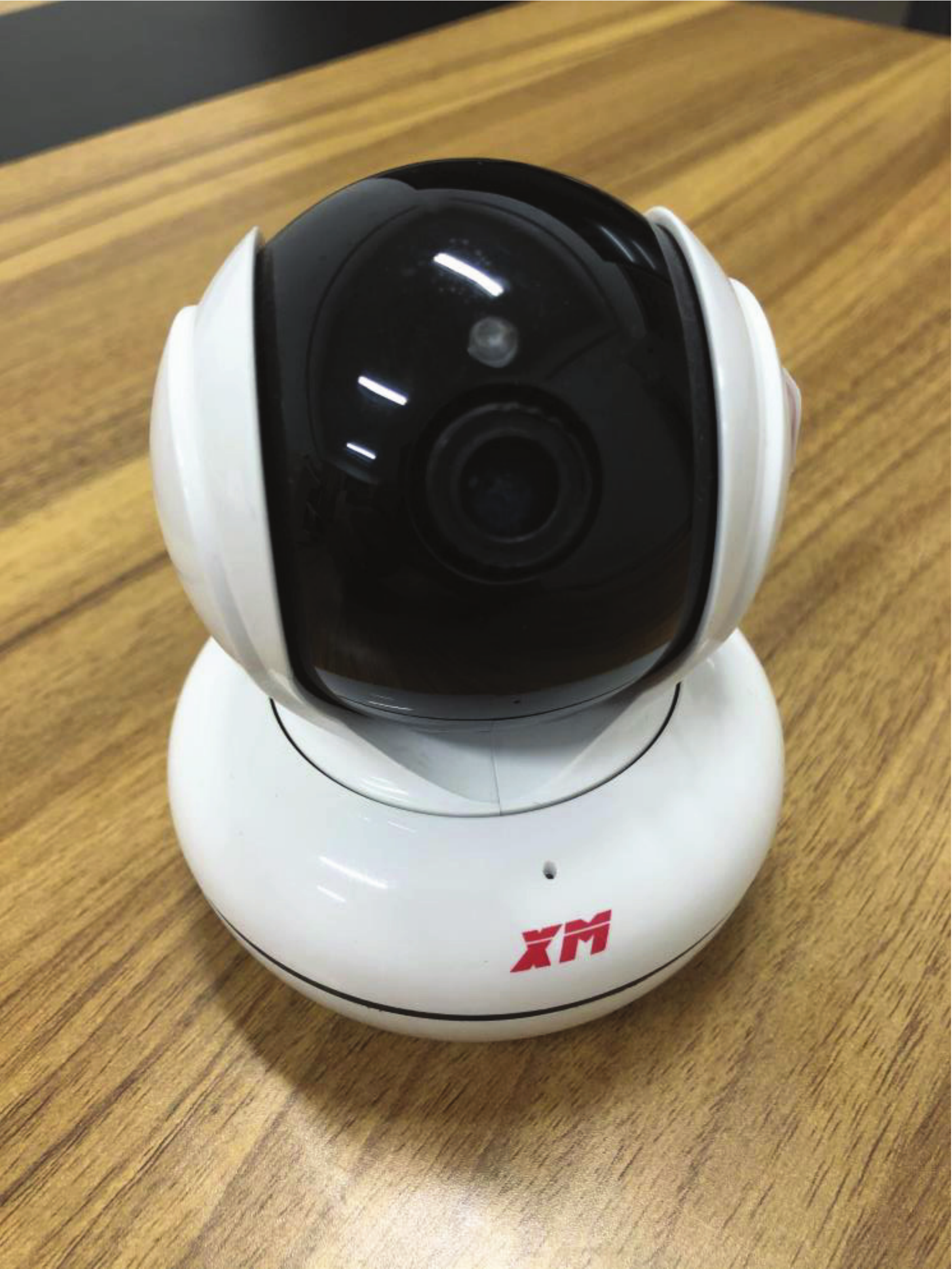}%
\end{minipage}
}

\subfigure[Edimax Camera]{%
\begin{minipage}[t]{0.21\linewidth}
\centering
\includegraphics[width=1.1\textwidth]{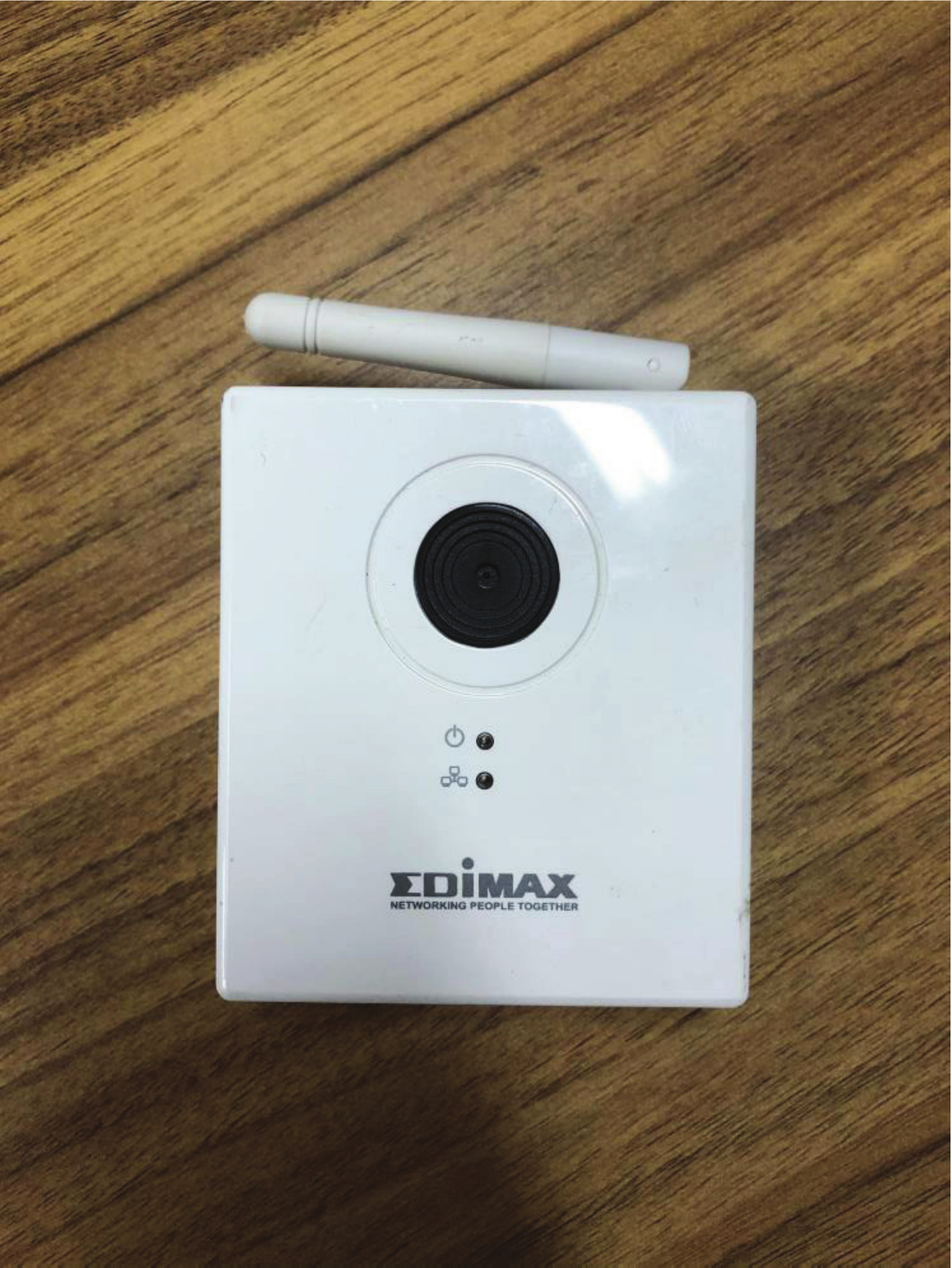}%
\end{minipage}
}
\subfigure[Edimax\textcolor{white}{++} Plug]{%
\begin{minipage}[t]{0.21\linewidth}
\centering
\includegraphics[width=1.1\textwidth]{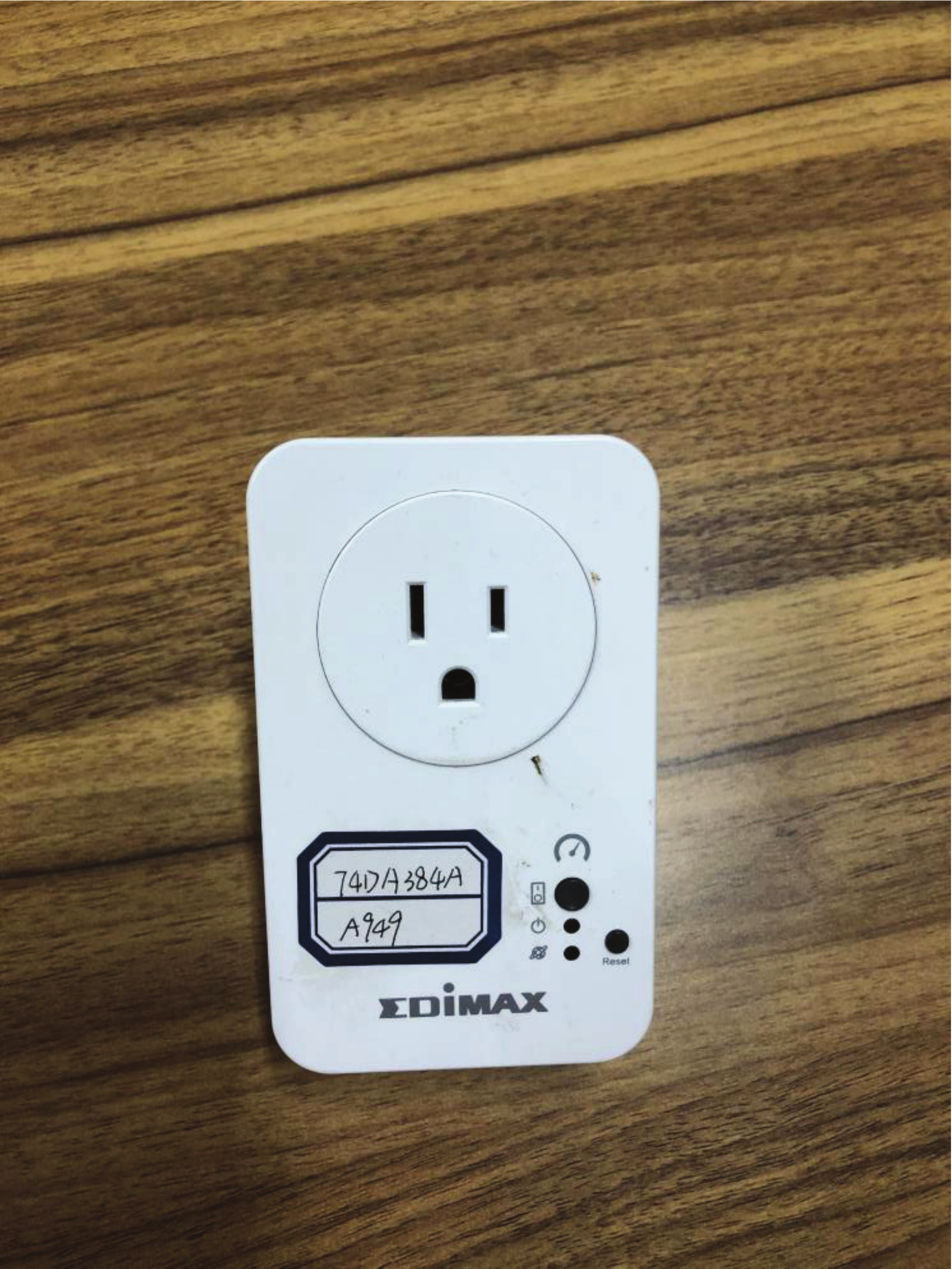}%
\end{minipage}
}
\subfigure[PurpleAir Sensor]{%
\begin{minipage}[t]{0.21\linewidth}
\centering
\includegraphics[width=1.1\textwidth]{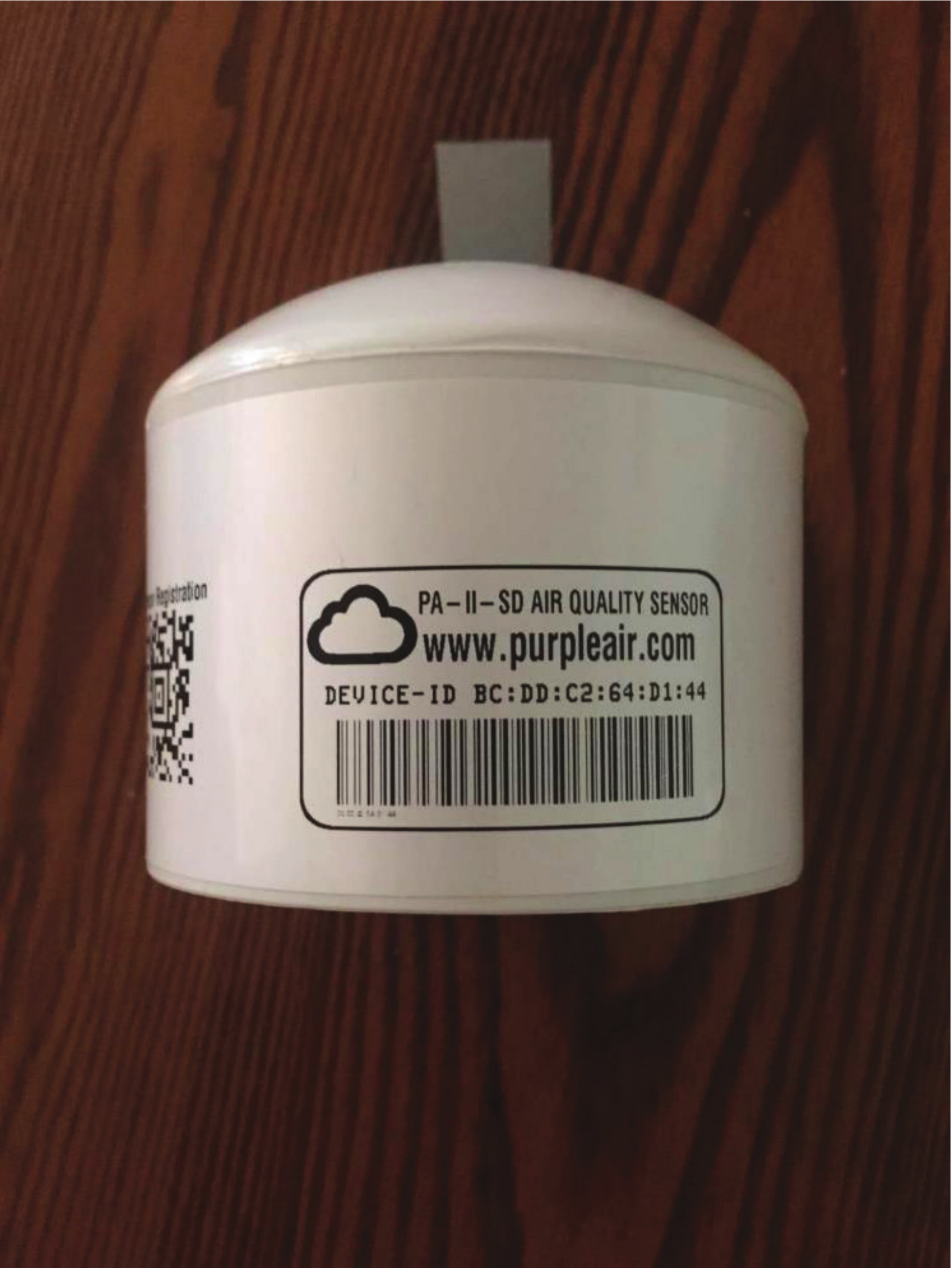}%
\end{minipage}
}

\centering
\caption{IoT devices analyzed with our framework}
\label{fig::devices}
\end{figure}

Fig. \ref{fig::devices} shows all devices we have reverse engineered, including  Edimax camera \cite{Globecom::ling::2017}, Edimax smart plug \cite{IoJT::ling::2017} and PurpleAir air quality monitoring sensor \cite{LZP+::AirQualitySensor::2018,GLZ+::MCU::2019} in our previous work. 
The PurpleAir air quality monitoring sensors are actually bare metal systems based on microcontrollers (MCUs) without an OS like Linux. Now our manual reverse engineering framework is still valid. Particularly OpenOCD and GDB can be used to debug the MCU firmware through JTAG.
We now briefly introduce how we used the framework to analyze the other devices that we are the first to have reverse engineered.

We reverse engineered the communication protocol of the D-Link cloud camera system.
The camera uses a read-only filesystem and we are able to find the CA certificate. As proposed in Section \ref{subsec:obtaining}, we replace the certificate by generating a new firmware with a forged root certificate and flash the new firmware into the target camera through the device management interface. Therefore, we can decrypt the TLS/SSL encrypted traffic, and finally find that the camera is also under the risk of spoofing attacks.

We reverse engineered the communication protocol of the Haier IP camera and the Xiongmai IP camera and find they are vulnerable to the spoofing attack and the Xiongmai IP camera also under an unauthorized access attack. 
(i) For the Haier IP camera, we find the app is packed to hide the executable files, i.e., dex files. To extract the dex files from the packed app \cite{ESORICS::dexhunter::2015}, we use Xposed and Fdex2 \cite{fdex2}, which is a module of Xposed, to hook the \textit{loadclass} function and extract the dex files. Then, we can hook the app with Xposed and Frida, and perform static data flow analysis and dynamic debugging to the binaries of IoT device using GDB to discover the communication protocol, as shown in Section \ref{subsec:protocolanalysis}.
(ii) For the Xiongmai IP camera, we diassemble the camera app for static analysis and use code instrumentation techniques such as hooking through Frida \cite{Frida:2020} to analyze the app side communication protocol. We also disassemble the firmware, embed gdbserver onto a flash and use GDB to dynamically debug the binary files of the firmware. 

\vspace{-0.1cm}
\subsection{Limitations}
Our communication protocol reverse engineering framework has the following limitations. If an IoT device employs secure boot and the firmware image verification key is in secure storage such as e-fuse, we may not be able to change the firmware of the device, since secure boot will detect the change and refuse to start the device. Similarly, if flash encryption is enabled and the related keys are in secure storage, we cannot change the device firmware since we cannot obtain these keys. However, we find few IoT products use such 
secure boot and flash encryption.

%% file: 7RelatedWork.tex
\vspace{-0.1cm}
\section{Related Work}
\label{sec::RelatedWork}

In this section, we review the existing technologies for analyzing the security of IoT devices and Android apps. Particularly, we divided the state of art into three categories, i.e., static analysis, dynamic analysis, and hybrid analysis approaches.

\textbf{Static analysis}: Some static analysis approaches have been proposed to analyze the security of the IoT device firmware \cite{Security::costin::2014, Globecom::ling::2017, ASPLOS::David::2018, CAV::Kinder::2008, Infocommunications::papp::2019, iot-j::Shwartz::2018} and Android apps \cite{gallingani2014static, NDSS::lu::2015, NDSS::gordon::2015, nirumand2019vandroid, WASA::biswas::2017, JNCA::Seo::2014}. For example, Costin et al. \cite{Security::costin::2014} preformed a large-scale static analysis of IoT device firmware with correlation engine which could evaluate the similarity between the target IoT device firmware and the vulnerable ones so as to determine whether the target firmware contains existing vulnerabilities. Nirumand et al. \cite{nirumand2019vandroid} proposed a MDRE (Model Driven Reverse Engineering) based static analysis method to discover the security risks in the Android app communication. The static analysis approaches are fast and can reach comprehensive code coverage of the firmware or app \cite{COMPSAC::Aggarwal::2006, IWCMC::zheng::2014}. However, some IoT device firmware and Android apps are obfuscated or encrypted which cannot be disassembled and statically analyzed \cite{Mobisys::Hao::2014, PST::martinelli::2016}. In addition, the runtime behavior such as user input could not be statically determined and static analysis may cause false positives and false negatives \cite{NDSS::SMV::2014, IWCMC::zheng::2014, COMPSAC::Aggarwal::2006}.

\textbf{Dynamic analysis: }Dynamic analysis approaches could observe the runtime behavior of the target app and IoT device firmware and could be used to verify the correctness of the results of static analysis approaches by running the app or IoT device firmware with test cases \cite{COMPSAC::Aggarwal::2006}. For Android apps, Zheng et al. \cite{IWCMC::zheng::2014} proposed a dynamic analysis framework based on ptrace (process trace) which is a system call that could be used by one process to control another. The framework uses ptrace to monitor selected system calls to dynamically analyze malicious behaviors of the binary. The frameworks of dynamic analysis methods for IoT device firmware can be divided into two categories, i.e., software emulator based frameworks as well as the real IoT device hardware and the emulator based frameworks. For the first category, the IoT device firmware is performed on a software emulator and applied the dynamic analysis methods \cite{NDSS::chen::2016, RAID::gustafson::2019, Security::feng::2020}. For example, Chen et al. \cite{NDSS::chen::2016} presented FIRMADYNE, which is a dynamic debugging framework based on the emulator with an instrumented kernel. 14 previously-unknown vulnerabilities were discovered by using FIRMADYNE with automated webpages analysis and manual analysis.

Since the IoT device hardware is fairly diverse, it is nontrivial to emulate various IoT device hardware with software emulators \cite{NDSS::zaddach::2014}. To address this problem, some frameworks have been proposed, which relay I/O accesses between real IoT device hardware and the emulator \cite{NDSS::zaddach::2014, BAR::marius::2018, Security::corteggiani::2018}. For instance, Zaddach et al. \cite{NDSS::zaddach::2014} presented Avatar, which is a framework that dynamically analyzes the IoT devices by combining the emulator and the real hardware. The framework forwards the I/O accesses from the emulator to the real IoT device. The framework was evaluated with KLEE symbolic execution engine and existing fuzzing tools. However, dynamic analysis is time-consuming as it requires numerous test cases to ensure a certain degree of credibility for vulnerability detection. In addition, it is difficult to generate valid test cases \cite{NDSS::SMV::2014, COMPSAC::Aggarwal::2006}.

\textbf{Hybrid analysis: }Hybrid analysis methods, which combines the static and dynamic analysis technologies, have been proposed \cite{PST::martinelli::2016, IJIS::Spreitzenbarth::2015, Globecom::wang::2015, NDSS::SMV::2014, ICIoT::Geancarlo::2017, JCSSE::Visoottiviseth::2018, ESORICS::yao::2019, COMPSAC::Aggarwal::2006} to improve the accuracy of vulnerability discovery. For example, Martinelli et al. \cite{PST::martinelli::2016} proposed a framework to detect malicious apps by performing both static and dynamic analyzing approaches. They evaluated the framework using 2794 malicious apps with high detection accuracy. Palavicini et al. \cite{ICIoT::Geancarlo::2017} performed static analysis on IoT firmware to avoid path explosion when dynamically analyzing complex binaries with symbolic execution using a software emulator. Yao et al. \cite{ESORICS::yao::2019} identified a previously unknown vulnerability which is known as privilege separation vulnerability. They leveraged firmware loading information extraction, library function recognition, and symbolic execution methods to analyze the IoT device firmware and located 69 of 106 firmware containing privilege separation vulnerabilities.

Those existing technologies could not be used to probe into the various vulnerabilities located in the communication protocol of IoT systems \cite{Infocommunications::papp::2019, iot-j::Shwartz::2018, DSN::Chen::2019, Security::zhou::2019} and there is little systematically communication protocol reverse engineering approaches, since it is a great challenge to reverse engineer these protocols given the diversity of protocol implementation. For example, Papp et al. \cite{Infocommunications::papp::2019} and Shwartz et al. \cite{iot-j::Shwartz::2018} proposed the methods for reverse engineering IoT devices. They only focus on discovering the vulnerabilities in the firmware of IoT device 
instead of the security analysis of the communication protocol between the controller and device. To tackle this problem, we propose a framework to reverse engineering communication protocols of Linux based IoT systems for further protocol security analysis in this paper.

%% file: 8Conclusion.tex
\vspace{-0.1cm}
\section{Conclusion}
\label{sec::Conclusion}

In this paper, we propose a framework to manually reverse engineer the communication protocols of IoT devices so that the discovered protocol can be used for further security analysis. The framework works as follows: obtaining the app and firmware of an IoT device, collecting network traffic generated by the device and control app, defeating traffic protection, and discovering the communication protocol through traffic analysis, static analysis and dynamic analysis of the app and firmware. We present a case study of using the framework to reverse engineer the communication protocols of the WeMo smart plug. Once the plug's communication protocols are discovered, we are able to identify a crucial authentication vulnerability that allows the plug sharing attack to control victim plugs and connection hijacking attack for DoS. We demonstrate our framework is generic and could be applied to a variety of embedded Linux based IoT systems using either read-only or writable filesystems. We also briefly discussed how we applied the framework to a few other real-world IoT products and systems. We are the first to systematically propose such a manual communication protocol reverse engineering framework.